\documentstyle[12pt,epsf]{article} 
\begin{document}
\baselineskip 14pt plus 2pt minus 2pt
\newcommand{\beq}{\begin{equation}}
\newcommand{\eeq}{\end{equation}}
\newcommand{\beqa}{\begin{eqnarray}}
\newcommand{\eeqa}{\end{eqnarray}}
\newcommand{\dfrac}{\displaystyle \frac}
\renewcommand{\thefootnote}{\#\arabic{footnote}}
\newcommand{\ve}{\varepsilon}
\newcommand{\krig}[1]{\stackrel{\circ}{#1}}
\newcommand{\barr}[1]{\not\mathrel #1}
\newcommand{\vs}{\vspace{-0.3cm}}
\begin{titlepage}

\hfill hep--ph/9711365

\begin{center}

\vspace{2.0cm}

%{\large  \bf { CHIRAL STRUCTURE OF THE NUCLEON}}
%{\large  \bf { CHIRAL  NUCLEON DYNAMICS}}
{\Large  \bf {Chiral nucleon dynamics}}

\vspace{1.2cm}
                              
{\large  Ulf-G. Mei\ss ner}

\vspace{0.7cm}

Forschungszentrum J\"ulich, Institut f\"ur  Kernphysik
(Theorie)\\ D--52425 J\"ulich, Germany

\vspace{0.5cm}

email: Ulf-G.Meissner@fz-juelich.de

\end{center}

\vspace{2.5cm}

\begin{center}

 ABSTRACT

\end{center}

\vspace{0.1cm}

\noindent
These lectures give an introduction to baryon chiral perturbation
theory. I show in detail how to construct the chiral effective
pion--nucleon Lagrangian in  the one loop approximation. 
Particular emphasis is put on 
the physics related to electromagnetic probes, as manifest in pion 
photo-- and electroproduction, Compton scattering off nucleons and
the electroweak form factors of the nucleon.
 Other topics discussed in
some detail are the meaning of low--energy theorems, pion--nucleon scattering,
the reaction $\pi N \to \pi \pi N$ and
isospin violation  in the pion--nucleon system.
The chiral predictions are confronted with the existing data. 
Some remaining problems and new developments are outlined.

\vspace{3.5cm}

\begin{center}

Lectures given at The Hampton Graduate Studies at Jefferson Lab 
(HUGS 97)\\ Newport News, Virginia, USA, June 1997

\end{center}

\vspace{2cm}

\vfill

\noindent KFA-IKP(TH)-97-20 \hfill November 1997

\end{titlepage}

\tableofcontents

\vspace{1.5cm}

%%%%%%%%%%%%%%%%%%%%%%%%%%%%%%%%%%%%%%%%%%%%%%%%%%%%%%%%%%%%%%%%%%%%%
\section{Introduction}
\label{sec:intro}
\def\theequation{\arabic{section}.\arabic{equation}}
\setcounter{equation}{0}

Over the last few years, a tremendous amount of very precise data
probing the structure of the nucleon  at low energies has become
available. With the advent of CEBAF at Jefferson Lab, this data 
base can increase significantly if the low energy region is given
sufficient attention. These data encode
information about the nucleon in the non--perturbative regime, i.e. 
at typical momentum scales were straightforward perturbation theory 
in the running strong coupling constant $\alpha_S (Q^2)$ is no longer
applicable. At
present, there exist essentially two approaches to unravel the physics
behind the wealth of empirical information. On one hand, one uses
models which stress certain aspects of the strong interactions
(but tend to neglect or forget about others). The other
possibilitiy is to make use of the symmetries of the Standard Model (SM)
and formulate an effective field theory (EFT) to systematically  explore
the strictures of these symmetries. In the case of QCD and in the
sector of the three light quarks u, d and s we know that there exist
an approximate chiral symmetry which is spontaneously broken with the
appearance of eigth Goldstone bosons, the pions, kaons and the eta. These
pseudoscalar mesons are the lightest strongly interacting particles
and their small but non--vanishing masses can be traced back to
the fact that the current masses $m_u$, $m_d$ and $m_s$ are small
compared to any typical hadronic scale, like e.g. the proton mass.
In the meson sector, the EFT is called chiral perturbation theory and
is well developed and applied  successfully to many reactions
\cite{GL1} \cite{GL2} (for  reviews, 
see e.g.\cite{UGM}\cite{eckrev}\cite{bkmrev}).  Of course,
there is also the lattice formulation of QCD which over the years has
shown great progress but is not yet in the status to discuss in detail
what will be mostly the topic here, namely the baryon structure as
accurately probed with real or virtual photons. Clearly, the baryons
(and in particular the nucleons I will mostly focus on) are not related
directly to the spontaneous chiral symmetry breakdown. However, their
interactions with pions and among themselves are strongly constrained
by chiral symmetry. This is, of course, known since the sixties (see
e.g. the lectures by Coleman \cite{col} and references
therein). However, to go beyond the current algebra or tree level
calculations, one needs a systematic power counting scheme as it was
first worked out for the meson sector by Weinberg \cite{wein79}. As
shown by Gasser et al. \cite{GSS}, the straightforward generalization
to the baryon sector leads to problems related to the non--vanishing
mass of the baryons in the chiral limit, i.e. one has an extra large
mass scale in the problem. Stated differently, baryon four--momenta
are never small compared to the chiral symmetry breaking scale
$\Lambda_\chi$, $m_B / \Lambda_\chi \sim 1$. This can be overcome by a
clever choice of velocity--dependent fields \cite{JM} which allows to
transform the baryon mass term in a string of $1/m_B$ suppressed
interaction vertices. Then, a consistent power counting scheme emerges
where the expansion in small momenta and quark masses can be mapped
one--to--one on a (Goldstone boson) loop expansion. 
A nice introduction to some of the topics, in particular of the formal
aspects, has been given by Ecker~\cite{ecker}. His lectures should be
consulted for some details not covered here.

These lectures will be organized as follows. After a short review of
the construction of the EFT in the presence of matter fields, I will 
give a sample calculation (of the pion cloud contribution to the
nucleon mass). I elaborate in some detail on the structure of the 
next--to--leading order terms of dimension two and three in the effective
Lagrangian. In particular, I show how certain operators with {\it
 fixed} coefficients arise in the heavy fermion formalism.
This is followed by a study of the
numerical values of the low--energy constants appearing at second
order and their phenomenological interpretation in terms of resonance
saturation. Consequently, I construct the complete dimension three
effective Lagrangian.  
Then, I will discuss the status of the low--energy
theorems in pion photo-- and electroproduction and Compton scattering.
In the following sections, chiral predictions are confronted with the
data. Some new developments and open problems are touched upon in 
section~\ref{sec:final}.

%%%%%%%%%%%%%%%%%%%%%%%%%%%%%%%%%%%%%%%%%%%%%%%%%%%%%%%%%%%%%%%%%%%%%
\section{Chiral perturbation theory with baryons}
\label{sec:CHPT}
\setcounter{equation}{0}

Chiral perturbation theory (CHPT) is the EFT of the SM at low
energies in the hadronic sector. Since as an EFT it contains all terms allowed
by the symmetries of the underlying theory \cite{wein79}, it should be viewed 
as a direct consequence of the SM itself. The two main assumptions
underlying CHPT are that 
\begin{enumerate}
\item[(i)] the masses of the light quarks u, d (and possibly s) can be
treated as perturbations (i.e., they are small compared to a typical 
hadronic scale of 1 GeV) and  that 
\item[(ii)] in the limit of zero quark masses, the chiral symmetry is
spontaneously broken  to its vectorial subgroup. The resulting Goldstone
bosons are the pseudoscalar mesons (pions, kaons and eta).
\end{enumerate}

\noindent CHPT is a systematic low--energy expansion around the 
chiral limit \cite{wein79} \cite{GL1} \cite{GL2} \cite{leut}. It is a
well--defined quantum field theory although it has to be renormalized
order by order. Beyond leading order, one has to include loop diagrams to
 restore unitarity perturbatively. Furthermore, Green functions calculated in
CHPT at a given order contain certain parameters that are not constrained by 
the symmetries, the so--called low--energy constants (LECs).
At each order in the chiral expansion, those LECs have to be
determined  from phenomenology (or can be estimated with some
model dependent assumptions). 
For a review of the wide field of applications of CHPT, 
see, e.g., Ref.~\cite{UGM}. 

In the baryon sector, a
complication arises from the fact that the baryon mass $m_B$ does not vanish
in the chiral limit \cite{GSS}. Stated differently, only baryon three--momenta
can be small compared to the hadronic scale. To see this in more
detail, let us consider the two--flavor case with $m_u = m_d = \hat
m$ and collect the proton and the neutron in a bi-spinor
\beq
\Psi  = \left( \begin{array}{c} 
                 p \\ n
\end{array} \right)  \quad ,
\label{psi}
\eeq
which transforms non--linearly under chiral transformations,
\beq
\Psi \to \Psi' = K(L,R,U(x)) \, \Psi \quad . 
\eeq 
Here, $K(L,R,U(x))$ is a non--linear function of the meson fields
collected in $U(x)$ and of $L,R \in$ SU(2) \cite{ccwz}
\cite{weinno}. It is defined via
\beq
u' = L u K^\dagger = K u R^\dagger 
 \, , \, {\rm with } \, \, U = u^2 \, , \, u'^2 = U'^2 \quad . \eeq
The unimodular unitary matrix $U$ ($U^\dagger U = U U^\dagger =1$,
det$(U)$ = 1) transforms linearly under chiral transformations,
\beq U \to U' = L U R^\dagger \quad . \eeq
It is most convenient to parametrize $U$ as follows
\beq \label{siggauge}
U = (\sigma + i \vec \tau \cdot \vec \pi \, ) / F \, , \quad
\sigma^2 + {\vec \pi \,}^2 = F^2 \, \, , \eeq
with $F$ the pion decay constant in the chiral limit, $F_\pi = F[1
+{\cal O}(\hat m)] =92.4$ MeV. It is now
straightforward to write down chiral covariant derivatives and
construct the lowest order effective Lagrangian
\beq {\cal L}_{\rm eff}   = {\cal L}_{\pi N}^{(1)} +
 {\cal L}_{\pi \pi}^{(2)} \eeq
\beq {\cal L}_{\pi N}^{(1)}  = \bar{\Psi}
 \left( i \gamma_\mu D^\mu - \krig{m} + \frac{1}{2}\krig{g}_A \gamma^\mu
   \gamma_5 u_\mu \right) \Psi \label{lpin} \eeq
\beq\label{Lpipi}  {\cal L}_{\pi \pi}^{(2)} =
\frac{F^2}{4} \langle  \nabla_\mu U \nabla^\mu U^\dagger
\rangle + \frac{F^2 M^2}{4} \langle U + U^\dagger \rangle
 \quad , \eeq
with $u_\mu = i u^\dagger \nabla_\mu U u^\dagger$ and $\langle \ldots
\rangle$ denotes the trace in flavor space.
Here, the superscript '$\circ$' denotes quantities in the chiral limit, i.e.
$Q = ~\stackrel{\circ}{Q} [ 1 + {\cal O}(\hat m)]$ 
(with the exception of $M$ which is
the leading term in the quark mass expansion of the pion mass and $F$),
$g_A$ is the axial--vector
coupling constant measured in neutron $\beta$--decay, $g_A =
1.26$ and $m$ denotes the nucleon mass. 
For the three flavor case, one has of course two axial
couplings. The chiral dimension (power) of the respective terms is
denoted by the superscripts '(i)' (i=1,2). The pertinent covariant
derivatives are 
\beq \nabla_\mu U = \partial_\mu U - i e A_\mu [Q,U] \eeq
\beq D_\mu \Psi = \partial_\mu \Psi + \frac{1}{2} \lbrace u^\dagger
\left( \partial_\mu -i e A_\mu Q \right)u + u
\left( \partial_\mu -i e A_\mu Q \right)u^\dagger \rbrace \Psi 
= \partial_\mu \Psi + \Gamma_\mu \Psi 
\label{conn} \eeq
with $Q= {\rm diag}(1,0)$ the (nucleon) charge matrix and I only consider
external vector fields, i.e. the photon $A_\mu$. $D_\mu$ transforms
homogeneously under chiral transformation, $D_\mu' = K D_\mu
K^\dagger$, and  $\Gamma_\mu$ is the connection \cite{ccwz}.
 To show the strength of the effective Lagrangian approach,
let me quickly derive the so--called Goldberger--Treiman relation
(GTR) \cite{GTR}. For that, I set $A_\mu = 0$ and expand the
pion--nucleon Lagrangian to order $\vec \pi$,
\beq {\cal L}_{\pi N}^{(1)}  = \bar{\Psi}
 \left( i \gamma_\mu \partial^\mu - \krig{m} \right) \Psi
 - \frac{\stackrel{\circ}{g_A}}{F} \bar \Psi \gamma^\mu
   \gamma_5 \frac{\vec \tau}{2} \Psi \cdot \partial_\mu \vec \pi 
+ \ldots \eeq
from which we read off the $NN\pi$ vertex in momentum space
\beq V_{NN\pi} = \frac{\krig{g}_A}{2F} \gamma_\mu q^\mu \gamma_5 \vec \tau
\quad , \eeq
where the momentum $q_\mu$ is out--going.
The transition amplitude for single pion emission off a nucleon takes the form
\beq T_{NN\pi} = -i\bar u (p') V_{NN\pi} u(p) = i
\frac{\krig{g}_A}{F} 
\krig{m} \bar u (p') \gamma_5 u(p) \tau^i \, \, , \label{gtr1} \eeq
where I have used the Dirac equation $\bar u (p') \gamma_\mu q^\mu 
\gamma_5 u(p) = -2 \krig{m} \bar u (p') \gamma_5 u(p)$ with $q
= p'-p$. Comparing eq.(\ref{gtr1}) with the canonical form of the
transition amplitude
\beq  T_{NN\pi} = i \krig{g}_{\pi N} \, \bar u (p') \gamma_5 u(p) \eeq
one arrives directly at the GTR,
\beq  \label{gtr}
\krig{g}_{\pi N} = \frac{\krig{g}_A  \krig{m}}{F} \, \, , \eeq
which is fulfilled within a few \% in nature. This relation is
particularly intriguing because it links the strong pion--nucleon
coupling constant to some weak interaction quantities like $g_A$ and $F_\pi$
as a consequence of the chiral symmetry. I will discuss this relation and the
deviation thereof in
more detail in section~11.6. Finally, if one wants to
discuss processes with  two (or more) nucleons in the initial and
final state, one has to add a string of terms of the type
\beq {\cal L}_{\bar \Psi \Psi \bar \Psi \Psi} +  
{\cal L}_{\bar \Psi \Psi \bar \Psi \Psi \bar \Psi \Psi} + \dots \eeq
which are also subject to a chiral expansion and contain LECs which
can only be determined in few or many nucleon (baryon) processes.

Clearly, the appearance of the mass scale $\krig{m}$ in eq.(\ref{lpin}) causes
trouble. To be precise, if one calculates the self--energy of the
nucleon mass to one loop, one encounters terms of dimension {\it
  zero}, i.e. in dimensional regularization one finds a term of the
type \cite{GSS}
\beq {\cal L}_{\pi N}^{(0)} =
c_0 \, \bar \Psi \Psi \, , \quad \, c_0 \sim \left(\frac{\krig{m}}{F}\right)^2
\frac{1}{d-4} + \ldots \quad , \label{massdiv} \eeq 
where the ellipsis stands for terms which are finite as $d \to 4$.
Such terms clearly make it difficult to organize the chiral expansion
in a straightforward and simple manner. They can only be avoided if
the additional mass scale $\krig{m} \sim 1$ GeV can be eliminated from the
lowest order effective Lagrangian. (Notice here the difference to the
pion case - there the mass vanishes as the quark masses are sent to
zero.) To do that, consider the mass of the nucleon large compared to
the typical external momenta transferred by pions or photons and write
the nucleon four--momentum as
\beq p_\mu = m \, v_\mu + \ell_\mu \, , \quad p^2 = m^2 \, , \quad v \cdot
\ell \ll m \, . \eeq
Notice that to this order we do not have to differentiate between $m$
and $\krig{m}$ and $v_\mu$ is the nucleon four--velocity (in the
rest--frame, we have $v_\mu =( 1 , \vec 0 \, )$). In that
case, we can decompose the wavefunction $\Psi$ into velocity
eigenstates \cite{JM} \cite{BKKM} 
\beq \Psi (x) = \exp [ -i \krig{m} v \cdot x ] \, [ H(x) + h(x) ] \eeq
with 
\beq \barr v \, H = H \,\, ,  \barr  v \, h = -h \,\, , \eeq
or in terms of velocity projection operators
\beq 
P_v^+ H = H \, , \, P_v^- h = h \, , \quad P_v^\pm =
\frac{1}{2}(1 \pm \barr v \,) \, , \quad P_v^+ + P_v^- = 1\,  . 
\eeq
One now eliminates the 'small' component $h(x)$ either by using the
equations of motion or path--integral methods.
The Dirac equation for the velocity--dependent
baryon field $H = H_v$ (I will always suppress the label '$v$') 
takes the form $i v \cdot \partial H_v = 0$ to lowest
order in $1/m$. This allows for a consistent chiral counting as described
below and the effective pion--nucleon Lagrangian takes the form:  
\beq {\cal L}_{\pi N}^{(1)}  = \bar{H}
 \left( i v \cdot D  + \krig{g}_A S \cdot u \right) H 
+ {\cal O}\left(\frac{1}{m} \right) \, , \label{lagr} \eeq
with $S_\mu$ the covariant spin--operator
\beq S_\mu = \frac{i}{2} \gamma_5 \sigma_{\mu \nu} v^\nu \, , \, 
S \cdot v = 0 \, , \, \lbrace S_\mu , S_\nu \rbrace = \frac{1}{2} \left(
v_\mu v_\nu - g_{\mu \nu} \right) \, , \, [S_\mu , S_\nu] = i
\epsilon_{\mu \nu \gamma \delta} v^\gamma S^\delta \,
 \, , \label{spin}\eeq
in the convention $\epsilon^{0123} = -1$. There is one subtlety to be
discussed here. In the calculation of loop graphs, divergences appear
and one needs to regularize and renormalize these. That is done most
easily in dimensional regularization since it naturally preserves the
underlying symmetries. However, the totally antisymmetric Levi-Civita
tensor is ill-defined in $d \neq 4$ space--time dimensions. One
therefore has to be careful with the spin algebra. In essence,
one has to give a prescription to uniquely fix the finite pieces.
The mostly used convention to do this is to only use the
anticommutator to simplify products of spin matrices and only
take into account that the commutator is antisymmetric under
interchange of the indices. Furthermore, $S^2$ can be uniquely
extended to $d$ dimensions via $S^2 = (1-d)/4$. With that in mind,
two important observations can be made. Eq.(\ref{lagr}) does not
contain the nucleon mass term any more and also, all Dirac matrices
can be expressed as combinations of $v_\mu$ and $S_\mu$ \cite{JM},
\begin{displaymath} \bar H \gamma_\mu H = v_\mu \bar H H \, , \, 
\bar H \gamma_5 H = 0 \, , \, \bar H \gamma_\mu \gamma_5 H =
2 \bar H S_\mu H \, \, ,  \end{displaymath}
\beq \label{spinga}
\bar H \sigma_{\mu \nu} H = 2 \epsilon_{\mu \nu \gamma
  \delta} v^\gamma \bar H S^\delta H \, , \, \bar H \gamma_5 
\sigma_{\mu \nu} H = 2i \bar H (v_\mu S_\nu - v_\nu S_\mu) H \, \, ,
\eeq
to leading order in $1/m$. More precisely, 
this means e.g. $\bar H \gamma_5 H = {\cal O}(1/m)$. 
 We read off the nucleon propagator,
\beq S_N (\omega ) = \frac{i}{\omega + i \eta} \, , \quad \omega = v
\cdot \ell \, , \quad \eta > 0\, \, , \label{prop} \eeq
and the Feynman insertion for the emission a pion with momentum $\ell$
from a nucleon is
\beq \frac{g_A}{F_\pi} \, \tau^a \, S \cdot \ell \quad . \label{vert}
\eeq
It is also instructive to consider the transition from the relativistic
fermion propagator to its counterpart in the heavy fermion limit.
Starting with $S_N (p) = i / (\barr{p} -m)$, one can project out
the light field component,
\beqa
S_N &=& P_v^+ \, i \, {\barr{p} -m \over p^2 -m^2}\, P_v^+ = i\,
{p \cdot v +m \over p^2 - m^2} \, P_v^+ = i \,
{2m + v \cdot \ell \over 2m  v \cdot \ell + \ell^2} \, P_v^+ \nonumber
\\
&& \simeq {i \over  v \cdot \ell + \ell^2/2m - ( v \cdot \ell)^2/2m
+ {\cal O}(1/m^2)} \, P_v^+ \simeq  {i\, P_v^+  \over  v \cdot \ell} +
 {\cal O}({1\over m}) \,\,\, ,
\eeqa
which in fact shows that one can include the kinetic energy
corrections to be discussed in more detail below already in the
propagator.
Notice that from now on I will  not always distinguish  between the
observables and their chiral limit values (although that distinction 
should be kept in mind). Before proceeding with some actual
calculations in heavy baryon CHPT (HBCHPT),
let me outline the chiral power counting which is used
to organize the various terms in the energy expansion. 

%%%%%%%%%%%%%%%%%%%%%%%%%%%%%%%%%%%%%%%%%%%%%%%%%%%%%%%%%%%%%%%%%%%%%%%
\section{Chiral power counting}
\label{sec:count}
\setcounter{equation}{0}

%%%power counting
To calculate any process to a given order, it is useful to have
a compact expression for the chiral power counting \cite{wein79} \cite{ecker}.
First, I will restrict myself to purely mesonic or single--baryon
processes. Since these arguments are general, I will consider the
three flavor case.
Any amplitude for a given physical process has a certain {\bf chiral} {\bf
dimension} $D$ which keeps track of the powers of external momenta and meson
masses (collectively labelled $d$). 
The building blocks to calculate this chiral dimension
from a general Feynman diagram in the CHPT loop expansion are 
(i) $I_M$ Goldstone boson (meson) propagators $\sim 1/(\ell^2 -M^2)$ 
(with $M=M_{\pi , K, \eta}$ the meson mass) of dimension $D= -2$, 
(ii) $I_B$ baryon propagators $\sim 1/ v \cdot \ell$ (in HBCHPT) with
$D= -1$, (iii) $N_d^M$ mesonic vertices with $d =2,4,6, \ldots$ and 
(iv) $N_d^{MB}$ meson--baryon vertices with $d = 1,2,3, \ldots$. 
Putting these together, the chiral dimension $D$ of a given amplitude reads
\begin{equation}
D =4L - 2I_M - I_B + \sum_d d( \,  N_d^M + N_d^{MB} \, )
%\label{e11}
\end{equation}
with $L$ the number of loops. For connected diagrams, one can use
the general topological relation
\begin{equation}
L = I_M + I_B -  \sum_d ( \,  N_d^M + N_d^{MB} \, ) + 1 
%\label{e12}
\end{equation}
to eliminate $I_M$~:
\begin{equation}
D =2L + 2 + I_B + \sum_d (d-2)  N_d^M + \sum_d (d-2) N_d^{MB} ~ .
\label{DLgen}
\end{equation}
Lorentz invariance and chiral symmetry demand $d \ge 2$ for mesonic 
interactions and thus the
term $\sum_d (d-2)  N_d^M$ is non--negative. Therefore, in the absence
of baryon fields, Eq.~(\ref{DLgen}) simplifies to \cite{wein79}
\begin{equation}
D =2L + 2 + \sum_d (d-2)  N_d^M  \, \ge 2L + 2 ~ .
\label{DLmeson}
\end{equation}
To lowest order $p^2$, one has to deal with tree diagrams 
($L=0$) only. Loops are suppressed by powers of $p^{2L}$. 

Another case of interest for us has a single baryon line running through 
the diagram (i.e., there is exactly one baryon in the in-- and one baryon 
in the out--state). In this case, the identity
\begin{equation}
\sum_d N_d^{MB} = I_B + 1 
%\label{e15}
\end{equation}
holds leading to \cite{ecker}
\begin{equation}
D =2L + 1  + \sum_d (d-2)  N_d^M + \sum_d (d-1)  N_d^{MB}  \, \ge 2L + 1 ~ .
\label{DLMB}
\end{equation}
Therefore, tree diagrams start to contribute at order $p$ and one--loop
graphs at order $p^3$. Obviously, the relations involving
baryons are only valid in HBCHPT.

Let me now consider diagrams with $N_\gamma$ external photons.
Since gauge fields like the electromagnetic field appear
in covariant derivatives, their chiral dimension is obviously $D=1$.
One therefore writes the chiral dimension of a general amplitude
with $N_\gamma$ photons as
\begin{equation}
D = D_L + N_\gamma  ~,
%\label{e17}
\end{equation}
where $D_L$ is the degree of homogeneity of the (Feynman) amplitude $A$ as 
a function of external momenta ($p$) and meson masses ($M$) in the
following sense (see also \cite{rho}):
\begin{equation}
A(p,M;C_i^r(\lambda),\lambda/M) 
= M^{D_L} \, A ( p/M , 1;C_i^r(\lambda), \lambda/M )  ~ ,
%\label{e18}
\end{equation}
where $\lambda$ is an arbitrary renormalization scale and $C_i^r(\lambda)$
denote renormalized LECs. From now on, I will suppress the explicit dependence
on the renormalization scale and on the LECs. Since the total amplitude
is independent of the arbitrary scale $\lambda$, one may in particular
choose $\lambda = M$.
Note that $A(p,M)$ has also a certain physical dimension (which is 
of course independent of the number of loops and is therefore in
general different from $D_L$). The correct
physical dimension is ensured by appropriate factors of $F_\pi$ and $m$
in the denominators.

Finally, consider a process with $E_n$ ($E_n = 4, 6, \ldots$) external
baryons (nucleons). The corresponding chiral dimension $D_n$ follows to be
\cite{weinnn}
\beq D_n = 2(L-C) + 4 - \frac{1}{2}E_n + \sum_i V_i \Delta_i \, ,
\quad \Delta_i = d_i + \frac{1}{2}n_i -2 \, \, , \eeq
where $C$ is  the number of connected pieces and one has $V_i$
vertices of type $i$ with $d_i$ derivatives and $n_i$ baryon fields
(these include the mesonic and meson-baryon vertices discussed before).
Chiral symmetry demands $\Delta_i \ge 0$. As before, loop diagrams
are suppressed by $p^{2L}$. Notice, however, that this chiral counting
only applies to the irreducible diagrams and not to the full S--matrix
since reducible diagrams can lead to IR pinch singularities and need
therefore a special treatment (for details, see refs.\cite{weinnn} 
\cite{weinnp}).

I will now briefly discuss the general structure of the effective
Lagrangian based on these power counting rules, restricting myself
again to the two--flavor case and processes with one nucleon in the
asymptotic in-- and out--states. While the lowest order Lagrangian 
eq.(\ref{lagr}) has $D=1$, one can construct a string of local
operators with $D=2,3,4, \ldots$. One--loop diagrams start at order
$p^3$ if one only uses insertions from ${\cal L}_{\pi
  N}^{(1)}$. Two--loop graphs are suppressed by two more powers of $p$
so that within the one--loop approximation one should consider tree
diagrams from
\beq
{\cal L}_{\pi N} = {\cal L}_{\pi N}^{(1)} + {\cal L}_{\pi N}^{(2)} +
{\cal L}_{\pi N}^{(3)} + {\cal L}_{\pi N}^{(4)}  \, \, , \label{lpin4} \eeq
and loop diagrams with insertions from ${\cal L}_{\pi N}^{(1,2)}$.
 It is important to stress that not all of the terms in 
${\cal L}_{\pi N}^{(2,3,4)}$ contain LECs, but some of  the
coefficients are indeed fixed for kinematical or similar reasons. I
will discuss the general structure of ${\cal L}_{\pi N}^{(2)}$ in
section~\ref{sec:lpin2}.
It should also be stressed that although there are many terms in 
${\cal L}_{\pi N}^{(2,3,4)}$, for a given process most of them do {\it
  not} contribute. As an example let me quote the order $q^4$
calculation of the nucleons' electromagnetic polarizabilities
\cite{bkms} which involves altogether four LECs from ${\cal L}_{\pi N}^{(2)}$
and four from ${\cal L}_{\pi N}^{(4)}$, a number which can certainly be
controlled. To all of this, one has of course to add the purely
mesonic Lagrangian ${\cal L}_{\pi \pi}^{(2)}+{\cal L}_{\pi \pi}^{(4)}$
\cite{GL1} \cite{GL2}.

%%%%%%%%%%%%%%%%%%%%%%%%%%%%%%%%%%%%%%%%%%%%%%%%%%%%%%%%%%%%%%%%%%%%%%%%%%%%
\section{A simple calculation}
\label{sec:calc}
\setcounter{equation}{0}

In this section, I will present a typical calculation, namely the
nucleon mass shift from the pion loop (in the one--loop
approximation). The full result can e.g. be found in refs.\cite{dobo} 
\cite{BKKM}. A similar sample calculation has been given in the
lectures by Jenkins and Manohar \cite{dobo}, but I will use another
method which is easier to generalize to processes with external photons. 

Consider the Feynman diagram where the nucleon emits a pion of momentum
$\ell$ and absorbs the same pion (which is incoming with momentum
$-\ell$). Using eqs.(\ref{prop},\ref{vert}) and the relativistic
propagator for the pion, the mass shift $\delta m$ is given by
\beq 
\delta m = i \frac{3g_A^2}{F_\pi^2} \int \frac{d^d \ell}{(2\pi)^d}
\frac{i}{\ell^2 - M_\pi^2 + i\eta} \,  \frac{i}{-v \cdot \ell + i \eta}
 S \cdot (- \ell) \,  S \cdot \ell \, , \eeq
making use of $\tau^a \tau^a = 3$. From the anti--commutation relation 
of two spin matrices, eq.(\ref{spin}), and by completing the square we have
\beq S_\mu S_\nu \ell^\mu \ell^\nu = \frac{1}{4} \left( v \cdot \ell
\, v \cdot \ell + M_\pi^2 - \ell^2 - M_\pi^2 \right) \, , \eeq
so that 
\beq 
\delta m = i \frac{3g_A^2}{4 F_\pi^2} \int \frac{d^d \ell}{(2\pi)^d} \left[
\frac{1}{v \cdot \ell -i \eta} +
\frac{v \cdot \ell}{M_\pi^2 - \ell^2 -i\eta} - \frac{M_\pi^2}{(M_\pi^2
  - \ell^2 -i\eta)(v \cdot \ell - i\eta)} \right] \, . \label{int} \eeq
To calculate this integral, we make use of dimensional
regularization. The first term in the square brackets vanishes in
dimensional regularization (see e.g. ref.\cite{coll}) and the second
one is odd under $\ell \to -\ell$, i.e. it also vanishes. So we are
left with
\beq \delta m =  \frac{3g_A^2}{4 F_\pi^2} \, J(0) \, M_\pi^2 \eeq
\beq J(0) = \frac{1}{i} \, \int \frac{d^d \ell}{(2\pi)^d}
\frac{1}{(M_\pi^2  - \ell^2 -i\eta)(v \cdot \ell - i\eta)} \quad . \eeq
The remaining task is to evaluate $J(0)$. For that, we use the
identity
\beq \frac{1}{AB} = \int_0^\infty dy \, \frac{2}{[A+2yB]^2} \, \, , \eeq
define $\ell' = \ell - y v$, complete the square and use $v^2 =1$,
\beq J(0)  = \frac{1}{i} \, \int_0^\infty dy \, \int \frac{d^d
\ell '}{(2\pi)^d} \, \frac{1}{[M_\pi^2 + y^2 + {\ell'}^2 - i \eta]^2}
\eeq
\beq \quad \quad \quad \quad \quad  = 
\frac{2}{(2\pi)^d} \, \int_0^\infty dy \, \int_0^\infty d\ell'
\frac{(\ell')^{d-1}}{[M_\pi^2 + y^2 + {\ell'}^2]^2} \, 
\frac{2 \pi^{d/2}}{\Gamma(d/2)} \, \, , \label{j0} \eeq
where we have performed a Wick rotation, 
$\ell_0 \to i \ell_0$ and dropped the $i
\eta$. The last factor in eq.(\ref{j0}) is the surface of the sphere
in $d$ dimensions. Introducing polar coordinates, $y = r \cos \phi$ and
$\ell' = r \sin \phi$ and noting that the Jacobian of this
transformation  is $r$, we have
\beq J(0) = \frac{4(4 \pi)^{-d/2}}{\Gamma(d/2)} \, \int_0^\infty dr \,
\frac{r^d}{(r^2+M_\pi^2)^2} \,  \int_0^{\pi/2} d\phi \, (\sin
\phi)^{d-1} \, . \label{ja} \eeq
We then perform the further substitution $r = M_\pi \tan \Phi$ in the 
$r$--integral. Then,
both integrals appearing in eq.(\ref{ja}) can be expressed in terms
of products of $\Gamma$ functions with the result
\beq J(0) = M_\pi^{d-3} \, (4 \pi)^{-d/2} \, \Gamma \left(
\frac{1}{2} \right) \, \Gamma \left(
\frac{3-d}{2} \right) = - \frac{M_\pi}{8 \pi} \, \, , \eeq
where in the last step I have set $d=4$. This leads us to
\beq \delta m  = -\frac{3g_A^2}{32 \pi} \, 
\frac{M_\pi^3}{F_\pi^2} \, \, . \label{self} \eeq
The pion loop leads to a self--energy $\Sigma$ which shifts the pole of
the nucleon propagator by $\delta m$, i.e.
\beq
m = \, \krig{m} + \Sigma (0) =  \, \krig{m} + \delta m \quad .
\eeq
%\beq \frac{i}{v\cdot \ell - i \Sigma} = \frac{i}{v \cdot \ell} \left(
%1 - \Sigma \frac{i}{v \cdot \ell} \right)^{-1} = \frac{i}{v\cdot
%  \ell}+ \frac{i}{v\cdot \ell} \, \Sigma \, \frac{i}{v\cdot \ell} + \ldots =
% \frac{i}{v \cdot \ell - \delta m} \, . \eeq
There are a few important remarks concerning eq.(\ref{self}). First,
the pion loop contribution is non--analytic in the quark masses
\cite{lapa}
\beq \delta m \sim (\hat m)^{3/2}    \, \, , \eeq
 since
\beq
M^2_\pi = 2 \, \hat m \, B \, [1+{\cal O}(m_{quark})]~.
\label{masses}
\eeq
The constant $B$ is related to the scalar quark condensate and is
assumed to be non--vanishing in the chiral limit (supported by lattice data),
$B = |\langle 0| \bar q q |0\rangle | / F^2$ and large, $B \sim 1.5\,$GeV. 
(For a different scenario, see e.g. refs.\cite{Stern}). Second, the
pion cloud contribution is attractive, i.e. it lowers the nucleon
mass, and third, it vanishes in the chiral limit, i.e it has the
expected chiral dimension of three (since it is a one--loop graph with
insertions from the lowest order effective Lagrangian). More detailed
studies of the baryon masses 
and $\sigma$--terms can be found in refs.\cite{gass} \cite{dobo}
\cite{liz} \cite{jms} \cite{bkmz} \cite{ll}. 
Here, I only wish to
state that at present it is not known whether the scalar three--flavour
sector (i.e. the baryon masses and the $\sigma$--terms) can be described
consistently within CHPT. The first complete calculation to order
$p^4$ has been presented in \cite{bora}, but the appearing LECs 
had to be modeled using resonance saturation including Goldstone boson
loops. I will later return to the questions surrounding the three
flavor sector and only remark here that much more work needs to be
done to achieve a consistent picture in the presence of kaon loops, 
the main  problem being that the
kaon loop corrections are rather large (since $(M_K / M_\pi)^3 \sim 46$)
and the corresponding LECs have therefore large coefficients to
compensate for the meson cloud contribution. It might, however, well be that
the corrections beyond leading order are much smaller. This can
only be decided upon a series of {\it complete} calculations to order $p^4$.

%%%%%%%%%%%%%%%%%%%%%%%%%%%%%%%%%%%%%%%%%%%%%%%%%%%%%%%%%%%%%%%%%%%%%%%%%%%
\section{The structure of ${\cal L}_{\pi N}^{(2)}$}
\label{sec:lpin2}
\setcounter{equation}{0}

In this section, I will first write down the order $p^2$ effective
Lagrangian, ${\cal L}_{\pi N}^{(2)}$, and then discuss some of its 
peculiarities. Allowing for the moment for $m_u \ne m_d$, its
most general form is (I only consider external scalar and vector
fields, the generalization to pseudoscalar and axial--vector ones is 
straightforward):
\begin{displaymath} 
{\cal L}_{\pi N}^{(2)} = \bar H \left\lbrace
\frac{1}{2\krig{m}}(v \cdot D)^2 - \frac{1}{2\krig{m}} D^2
-\frac{i \krig{g}_A}{2 \krig{m}} \lbrace S \cdot D , v \cdot u \rbrace
+ c_1 \, \langle \chi_+ \rangle 
+ \left(c_2 - \frac{\krig{g}_A^2}{8 \krig{m}} \right)
(v \cdot u)^2 \right. \end{displaymath}
\beq
 \left.
+ c_3 \, u \cdot u + \left( c_4 + \frac{1}{4\krig{m}} \right) [S^\mu,S^\nu]
u_\mu u_\nu +  c_5 \, \tilde{\chi}_+ -  \frac{i [S^\mu,S^\nu]}{4 \krig{m}} 
\left( (1 +c_6) f_{\mu \nu}^+ + c_7 \, {\rm Tr} \,  f_{\mu \nu}^+ \right)
 \right\rbrace H  \label{lp2}\eeq
with 
\beq \chi_\pm = u^\dagger \chi u^\dagger \pm u \chi^\dagger u \, , \quad
\tilde{\chi}_+ = \chi_+ - \frac{1}{2} \langle \chi_+ \rangle \, , \quad
f_{\mu \nu}^+ = u^\dagger F_{\mu \nu} u + u F_{\mu \nu} u^\dagger \, .
 \eeq
Here, $\chi = 2 B {\cal M}$ (${\cal M}$ is the quark mass matrix) and
$F_{\mu \nu} = \partial_\mu A_\nu - \partial_\nu A_\mu$ the canonical
photon field strength tensor. 
The term proportional to $c_5$ vanishes in the isospin limit $m_u= m_d$.
One could further reduce the number of terms by using the nucleon's
equation of motion (eom), $v\cdot D \, H = 0$ \cite{eckmoj}, 
but that is not necessary and sometimes complicates matters. In
what follows, I will always include  these eom terms in the effective
Lagrangian. One observes that some of the terms in eq.({\ref{lp2}) 
have no LECs but
rather fixed coefficients. The origin of this is clear, these terms
stem from the expansion of the relativistic $\pi N$ Lagrangian. To see
this in more detail, use the equation of motion for the small
component field $h(x)$,
\beq h = \frac{1}{2} (1 - \barr{v}) \frac{1}{2 \krig{m}} \left( i
\barr{D} +\frac{\krig{g}_A}{2} \barr{u} \gamma_5 \right) H + {\cal
  O}(1/ \krig{m}^2 )  \, \,  \eeq
to construct
\beq {\cal L}_{\pi N}^{(2)} = \frac{1}{2 \krig{m}} \bar H ( i
\barr{D} +\frac{\krig{g}_A}{2} \barr{u} \gamma_5 ) 
\frac{1 -  \barr{v}}{2} \, (i \barr{D} +\frac{\krig{g}_A}{2} \barr{u}
\gamma_5 ) H \, \, .  \label{lp2r} \eeq
Altogether, we have four different products of terms in eq.(\ref{lp2r}). Let me
consider the one proportional to $\barr{D} \barr{D}$, the other
contributions can be calculated in a similar fashion:
\beq \frac{i^2}{2 \krig{m}} \bar H \barr{D} \frac{1 -  \barr{v}}{2}
\barr{D} H = -\frac{1}{2 \krig{m}}D^\mu D^\nu \bar H \left[ \gamma_\mu
\frac{1 -  \barr{v}}{2} \gamma_\nu \right] H \, \, . \eeq
Straightforward application of the $\gamma$--matrix algebra allows us
to write the term in the square brackets on the r.h.s. as
\beq \bar H[ \dots ]H = \bar H [
g_{\mu \nu} - i \sigma_{\mu \nu} - \gamma_\mu v_\nu ] H \, \, , \eeq
where I have used that $P_v^+ H = H$. Collecting pieces, we have
\beq \frac{i^2}{2 \krig{m}} \bar H \left\lbrace  g_{\mu \nu} D^\mu
D^\nu  -  2 i \epsilon_{\mu \nu \alpha \beta}  D^\mu
D^\nu v^\alpha S^\beta  -  v_\mu D^\mu v_\nu D^\nu  \right\rbrace  H\, \,
. \eeq 
This simplifies further since $\epsilon_{\mu \nu \alpha \beta}$  and
$D^\mu D^\nu$ are anti-- and symmetric under $\mu \leftrightarrow
\nu$, respectively,
\beq \frac{1}{2 \krig{m}} \bar H \left\lbrace  -D^2 + (v \cdot D)^2 +
i \epsilon_{\mu \nu \alpha \beta} [D^\mu , D^\nu] v^\alpha S^\beta
\right \rbrace H \, \, . \eeq
Finally, the commutator of two chiral covariant derivatives is
related to the chiral connection $\Gamma_\mu$ via
\beq [ D_\mu , D_\nu] = \partial_\mu \Gamma_\nu - \partial_\nu
\Gamma_\mu +  [\Gamma_\mu , \Gamma_\nu ]  \, \, \, . \eeq
We can work this out for the explicit form of the connection given in 
eq.(\ref{conn}) and find
\beq \label{Dcomm} 
[ D_\mu , D_\nu] = -\frac{i}{2} f_{\mu \nu}^+ + \frac{1}{4}
[u_\mu , u_\nu] \quad . \eeq
Putting everything together, we find that the first three terms in
eq.(\ref{lp2}) plus the piece proportional to $\krig{g}_A^2 / (8
\krig{m})$ in the fifth and the piece proportional to $1/( 4 \krig{m})$ in
the seventh term are generated by expanding the operator $\barr{D} \barr{D}$.
 The first two contributions are corrections to the
kinetic energy and contain (besides others) a two--photon--nucleon
seagull which leads to the correct LET (low--energy theorem)
for low--energy Compton
scattering (see section \ref{sec:LET}). Similarly, the third term has
no free coefficient since it gives the leading term in the quark mass
expansion of the electric dipole amplitude $E_{0+}$ in neutral pion 
photoproduction off protons, $E_{0+} \sim M_\pi / m$. These terms have
no direct relativistic counter parts but are simply due to the
expansion of the relativistic pion--nucleon effective Lagrangian. 
Such constraints were
e.g. not accounted for in eq.(17) of ref.\cite{JM}. The finite
coefficients $c_{1,2,3,4}$ can be determined from the pion--nucleon
$\sigma$--term, S-- and P--wave $\pi N$ scattering lengths
and subthreshold parameters \cite{BKKM} \cite{bkmpin} \cite{bkmrev} 
\cite{bkmlec}. The last two terms in eq.(\ref{lp2})
are easy to pin down, they are related to the isoscalar and isovector
anomalous magnetic moment of the nucleon (in the chiral limit)
\cite{BKKM} \label{mmlec}
\beq c_6 = \krig{\kappa}_V \, \, \quad c_7 = \frac{1}{2} 
(\krig{\kappa}_S - \krig{\kappa}_V ) \quad . \eeq
Furthermore, the LEC $c_5$ can be determined from the strong contribution
to the neutron--proton mass difference.
The resulting values are summarized in table 1 (for more details,
please consult ref.\cite{bkmlec}) together with the results based on
resonance saturation as explained in the next section.
\renewcommand{\arraystretch}{1.3}
\begin{table}[hbt]
\begin{center}
\begin{tabular}{|l|r|r|r|c|}
   \hline
    $i$         & $c_i \quad \quad$   &  $c_i ' = 2m \, c_i$  &
                  $ c_i^{\rm Res} \,\,$ cv & 
                  $ c_i^{\rm Res} \,\,$ ranges    \\
    \hline
    1  &  $-0.93 \pm 0.10$  & $-1.74 \pm 0.19$ & $-0.9^*$ & -- \\
    2  &  $3.34  \pm 0.20$  & $6.27  \pm 0.38$ & $3.9\,\,$ & $2 \ldots 4$
 \\    
    3  &  $-5.29 \pm 0.25$  & $-9.92 \pm 0.47$& $-5.3\,\,$ 
                                     & $-4.5 \ldots -5.3$ \\
    4  &  $3.63  \pm 0.10$   & $6.81  \pm 0.19$ & $3.7\,\,$ 
                                     & $3.1 \ldots 3.7$ \\    
    5  &  $-0.09 \pm 0.01$   & $-0.17 \pm 0.02$ & $-\,\,$ & $-\,\,$ \\   
    \hline
    6           &  $5.83$             & --  & $6.1\,\,$  & $-\,\,$ \\
    7           &  $-2.98$            & --  & $-3.0\,\,$ & $-\,\,$ \\
    \hline
\end{tabular}
\end{center}
\caption{Values of the LECs $c_i$ in GeV$^{-1}$
and the dimensionless couplings $c_i '$
for $i=1,\ldots,5$. The LECs $c_{6,7}$ are dimensionless.
Also given are the central values (cv) and the ranges
for the $c_i$ from resonance
exchange as detailed in section~5. The $^*$ denotes an input
quantity.}
\end{table}

\medskip

The important lesson to be learned from this discussion is that in
HBCHPT we find in ${\cal L}_{\pi N}^{(2,3, \dots)}$ terms which have
no LECs but rather coefficients which are fixed. This is an artefact
of the {\it dual } expansion in small momenta $p$ versus the chiral
symmetry breaking scale {\it and} versus inverse powers
of the nucleon mass, i.e.
\beq \frac{p}{\Lambda_\chi } \, \, , \quad \frac{p}{m} \, \, . \eeq
In practice, since $m \sim \Lambda_\chi$, one has essentially one
expansion parameter besides the one due to the effect of the finite
quark masses. Since loops only appear at $D=3$, all the LECs in 
${\cal L}_{\pi N}^{(2)}$ are {\it finite} (as already should have
become clear from the previous discussion). It appears that the
numerical values of the dimension two LECs can be understood from 
resonance exchange as will be shown next.

%%%%%%%%%%%%%%%%%%%%%%%%%%%%%%%%%%%%%%%%%%%%%%%%%%%%%%%%%%%%%%%%%%%%%%%%%%%%%%%
\section{Phenomenological interpretation of the LECs}
\setcounter{equation}{0}

In this section, we will be concerned with the phenomenological
interpretation of the values for the LECs $c_i$. For that, guided by
experience from the meson sector \cite{reso}, we use resonance
exchange. Let me briefly elaborate on the Goldstone boson sector.
In the meson sector at next--to-leading order, the effective three flavor
Lagrangian ${\cal L}_{\phi}^{(4)}$ contains ten LECs, called $L_i$.
These have been determined from data in Ref.\cite{GL2}. 
The actual values of the
$L_i$ can be understood in terms of resonance exchange~\cite{reso}.
For that, one constructs the most general effective Lagrangian containing
besides the Goldstone bosons also resonance degrees of freedom.
Integrating out these heavy degrees of freedom from the EFT, one finds
that the renormalized $L_i^r (\mu=M_\rho)$ are practically saturated by 
resonance exchange ($S,P,V,A$). In some few cases, tensor mesons can
play a role~\cite{DT}. This is sometimes called {\it chiral
  duality} because part of the excitation spectrum of QCD reveals itself
in the values of the LECs. Furthermore, whenever vector and axial
resonances can contribute, the $L_i^r (M_\rho)$ are completely
dominated by $V$ and $A$ exchange, called {\it chiral} {\it VMD}~\cite{DRV}.
As an example, consider the finite (and thus scale--independent) LEC
$L_9$. Its empirical value is $L_9 = (7.1 \pm 0.3) \cdot 10^{-3}$. The
well--known $\rho$--meson (VMD) exchange model for the pion form
factor (neglecting the width), 
\beq F_\pi^V (q^2) = {M_\rho^2\over M_\rho^2 -q^2} = 1 + {q^2 \over M_\rho^2}
+ \ldots
\eeq
leads to
$L_9 = F_\pi^2 /(2 M_\rho^2) = 7.2 \cdot  10^{-3}$, by comparing with
the small momentum expansion of the pion form factor, $ F_\pi^V (q^2)
= 1 + \langle r^2 \rangle_\pi  \, q^2 / 6 + {\cal O}(q^4)$.
The resonance exchange result is
in good agreement with the empirical value.
Even in the symmetry breaking sector related to the quark mass, where
only scalar and (non-Goldstone) pseudoscalar mesons can contribute,
resonance exchange helps to understand why SU(3) breaking is generally
of ${\cal O}(25\%)$, except for the Goldstone boson masses.
Consider now an effective Lagrangian with
resonances chirally coupled to the nucleons and pions. One can
generate local pion--nucleon operators of higher dimension with given
LECs by letting the resonance masses become
very large with fixed ratios of coupling constants to masses,
symbolically
\beq  \tilde{{\cal L}}_{\rm eff}
 [U,M,N,N^\star] \to  {\cal L}_{\rm eff} [U,N]  \, \, , \eeq
That procedure amounts to decoupling the resonance degrees of freedom from
the effective field theory. However, the traces of these frozen
particles are encoded in the numerical values of certain LECs. In the
case at hand, we can have baryonic ($N^*$) and mesonic ($M$)
excitations,
\beq \label{cirdef}
c_i = \sum_{N^*=\Delta,R,\ldots} c_i^{N^*} + \sum_{M=S,V,\ldots} c_i^M
\,\,\, ,
\eeq
where $R$ denotes the Roper $N^* (1440)$ resonance. 
Consider first scalar ($S$) meson exchange. The SU(2) $S\pi\pi$
interaction can be written as
\beq
{\cal L}_{\pi S} = S \, \Big[ \bar{c}_m \, {\rm Tr}(\chi_+) + 
\bar{c}_d \, {\rm Tr} (u_\mu u^\mu) \Big] \,\,\, .
\eeq
{}From that, one easily calculates the s--channel scalar meson
contribution to the invariant amplitude $A(s,t,u)$ for elastic 
$\pi \pi$ scattering,
\beq
A^S (s,t,u) = \frac{4}{F_\pi^4 (M_S^2 -s)} \, [ 2 \bar{c}_m 
M_\pi^2 + \bar{c}_d (s- 2M_\pi^2) ]^2 +{16 \bar c_m M_\pi^2 \over 3
F_\pi^4 M_S^2 } \Big[ \bar c_m M_\pi^2 + \bar c_d(3s-4M_\pi^2) \big]   \,\,\, .
\eeq
Comparing with the SU(3) amplitude calculated in \cite{bkmsu3}, we
are able to relate the $\bar{c}_{m,d}$ to the  ${c}_{m,d}$ of
\cite{reso} (setting $M_{S_1} = M_{S_8} =M_S$ and using the large--$N_c$
relations $\tilde{c}_{m,d} = c_{m,d} / \sqrt{3}$ to express the
singlet couplings in terms of the octet ones),
$\bar{c}_{m,d} =   c_{m,d}/{\sqrt{2}}$,
with $|c_m| =42\,$MeV and $|c_d| = 32\,$MeV \cite{reso}. Assuming now
that $c_1$ is entirely due to scalar exchange, we get
\beq 
c_1^S = - \frac{g_S \, \bar{c}_{m}}{M_S^2} \quad .
\eeq
Here, $g_S$ is the coupling constant of the scalar--isoscalar meson to
the nucleons, ${\cal L}_{SN} = - g_S \, \bar{N} N \, S$. What this 
scalar--isoscalar meson is essentially doing is to mock up the strong
pionic correlations coupled to nucleons. Such a phenomenon 
is also observed in the meson sector. The one loop description
of the scalar pion form factor fails beyond energies of 400 MeV,
well below the typical scale of chiral symmetry breaking,
$\Lambda_\chi \simeq 1\,$GeV. Higher loop effects are needed
to bring the chiral expansion in agreement with the data \cite{game}.
Effectively, one can simulate these higher loop effects by introducing
a scalar meson with a mass of about 600 MeV. This is exactly the line
of reasoning underlying the arguments used here (for a pedagogical 
discussion on this topic, see \cite{cnpp}). It does, however, not mean
that the range of applicability of the effective field theory is
bounded by this mass in general. In certain channels with strong pionic
correlations one simply has to work harder than in the channels
where the pions interact weakly (as demonstrated in great detail
in \cite{game}) and go beyond the one loop approximation
which works well in most cases. For $c_1$ to be 
completely saturated by scalar exchange, $c_1 \equiv c_1^S$, we need
\beq
\frac{M_S}{\sqrt{g_S}} = 180 \, {\rm MeV} \quad .
\eeq
Here we made the assumption that such a scalar has the same couplings
to pseudoscalars as the real $a_0 (980)$ resonance.
It is interesting to note that the effective $\sigma$--meson in the 
Bonn one--boson--exchange potential \cite{bonn} with $M_S = 550\,$MeV and
$g_S^2/(4 \pi) = 7.1$ has $M_S / \sqrt{g_S} = 179\,$MeV. This number
is in stunning agreement with the the value demanded from scalar meson
saturation of the LEC $c_1$. With that, the scalar meson contribution
to $c_3$ is fixed including the sign, since $c_m c_d >0$ (see ref.\cite{reso}),
\beq
c_3^S = -2 \frac{g_s \, \bar{c}_d}{M_S^2} = 2 \frac{c_d}{c_m} \, c_1 
= -1.40 \, {\rm GeV}^{-1} \quad .
\eeq
The isovector $\rho$ meson only contributes to $c_4$. Taking a universal
$\rho$--hadron coupling and using the KSFR relation, we find
\beq
c_4^\rho = \frac{\kappa_\rho}{4m} = 1.63 \, {\rm GeV}^{-1} \quad ,
\eeq
using $\kappa_\rho = 6.1 \pm 0.4$ from the analysis of the nucleon
electromagnetic form factors, the process $\bar{N}N \to \pi \pi$
\cite{mmd} \cite{hoehpi} and the phenomenological one--boson--exchange
potential for the NN interaction.
I now turn to the baryon excitations. Here, the dominant one is the 
$\Delta (1232)$. Using the isobar model and the SU(4) coupling
constant relation (the dependence on the off--shell parameter $Z$ has 
already been discussed in \cite{bkmrev}), the $\Delta$ contribution to
the various LECs is readily evaluated,
\beq \label{delta}
c_2^\Delta = -c_3^\Delta = 2 c_4^\Delta = \frac{g_A^2 \, (m_\Delta
-m)}{2 [(m_\Delta -m)^2 - M_\pi^2]} = 3.83 \, {\rm GeV}^{-1} \,\, .
\eeq
These numbers are taken as the central values of the $\Delta$
contribution in what follows.
Unfortunately, there is some sizeable uncertainty related to these.
Dropping e.g. the factor $M_\pi^2$ in the denominator
of eq.(\ref{delta}), the numerical value decreases to 2.97 GeV$^{-1}$.
Furthermore, making use of the Rarita--Schwinger formalism and varying
the parameter $Z$, one can get sizeable changes in the $\Delta$
contributions ( e.g. $c_2^\Delta = 1.89, \, c_3^\Delta =-3.03 , 
\, c_4^\Delta = 1.42$ in GeV$^{-1}$ for $Z=-0.3$).
From this, we deduce the following ranges: $c_2^\Delta = 
1.9 \ldots3.8, \, c_3^\Delta =-3.8 \ldots -3.0 , 
\, c_4^\Delta = 1.4 \ldots 2.0$ (in GeV$^{-1}$).  
The Roper $N^* (1440)$ resonance contributes only marginally,
see ref.\cite{bkmlec}
Putting pieces together, we have for $c_2$, $c_3$ and $c_4$ from
resonance exchange (remember that $c_1$ was assumed to be saturated 
by scalar exchange)
\beqa \label{cireso}
c_2^{\rm Res} &=& c_2^\Delta + c_2^R =  3.83 + 0.05 = 3.88 \,\, , \nonumber \\
c_3^{\rm Res} &=& c_3^\Delta + c_3^S + c_3^R 
= -3.83 -1.40 - 0.06  = -5.29 \,\, , \nonumber \\
c_4^{\rm Res} &=& c_4^\Delta + c_4^\rho + c_4^R 
= 1.92 + 1.63 + 0.12 = 3.67 \,\, ,
\eeqa
with all numbers given in units of GeV$^{-1}$.
Comparison with the empirical values listed in table~1 shows that
these LECs can be understood  from resonance
saturation, assuming only that $c_1$ is entirely given by scalar meson
exchange. As argued before, the scalar meson parameters needed for
that are in good agreement with the ones derived from fitting NN
scattering data and deuteron properties  within the framework of a
one--boson--exchange model. We stress again that this $\sigma$--meson is an
effective degree of freedom which parametrizes the strong $\pi \pi$
 correlations (coupled to nucleons) in the scalar--isoscalar channel.
It should not be considered a novel degree of freedom which limits the
applicability of the effective field theory to a lower energy scale.
As pointed out before, there is some sizeable uncertainty related to the
$\Delta$ contribution as indicated by the ranges for the $c_i^{\rm Res}$
in table~1. It is, however, gratifying to observe that the empirical
values are covered by the band based on the resonance exchange model.
The LECs $\krig \kappa_s=-0.12$ and $\krig \kappa_v = 5.83$ can be estimated 
from neutral vector meson exchange using Eq.(\ref{mmlec}).
For the values from \cite{mmd}, $\kappa_\omega = -0.16 \pm
0.01$ and $\kappa_\rho = 6.1 \pm 0.4$, we see that the isoscalar and isovector
anomalous magnetic moments in the chiral limit can be well understood from 
$\omega$ and $\rho^0$ meson exchange. It is amusing that the isovector pion
cloud of the nucleon calculated to one loop allows to explain the observed 
difference between $\kappa_\rho$ and  $\kappa_v$. In strict vector meson
dominance these would be equal. It is well known \cite{hoeh} that the low 
energy part of the nucleon isovector spectral functions can not be 
understood in terms of the $\rho$--resonance alone (see also section~10).

%%%%%%%%%%%%%%%%%%%%%%%%%%%%%%%%%%%%%%%%%%%%%%%%%%%%%%%%%%%%%%%%%%%%%%%%%%%%%
\section{Construction and structure of ${\cal L}_{\pi N}^{(3)}$ }
\label{sec:lpin3}
\setcounter{equation}{0}

At the next order, $p^3$,
divergences appear. Ecker \cite{eckp3} has first calculated the
full determinant using heat--kernel methods and given the divergent
terms,
\beq {\cal L}_{\pi N}^{(3)} = \frac{1}{(4\pi)^2} \, \sum_{i=1}^{22} \,
b_i \, \bar H (x) \, O_i (x) \, H(x)       \eeq
with 
\beq
b_i = b_i^r (\lambda) + \Gamma_i \, L(\lambda) 
\, \, , \quad 
L = {\mu^{d-4}\over 16\pi^2} \biggl\{ {1 \over d-4} - {1\over 2}
 \biggl[ \ln(4\pi ) + \Gamma '(1) +1 \biggr] \biggr\}  
\quad . \eeq
The $O_i$ are monomials in the fields and have dimension three. Their
explicit forms together with the values of the $\Gamma_i$ can be found
in ref.\cite{eckp3}. Only a few of the finite $b_i^r$ have either been
determined  from phenomenology or estimated from resonance exchange
\cite{liz} \cite{BKKM} \cite{bkmpin} \cite{bkmpi0}. Furthermore,
Ecker and Moj\v zi\v s \cite{eckmoj} have constructed all finite terms
and used field redefinitions to find a minimal set of independent
operators at dimension three. I will now outline how one arrives at
this list and, in addition, show how one arrives at
the $1/m$ and $1/m^2$ corrections
to the various counterterms not listed in \cite{eckmoj}. Such
corrections can of course be absorbed in the numerical values of the
corresponding LECs, but for the reasons discussed before, I prefer to
keep them explicitely.  The construction of ${\cal L}_{\pi N}^{(3)}$ 
is done in five steps, which are: (1) enumeration of the building
blocks for relativistic spin--1/2 fields chirally coupled to pions and
external sources, (2) saturation of the free Lorentz indices by elements of the
underlying Clifford algebra and other operators of chiral dimension
zero (and one), (3) construction of the overcomplete relativistic Lagrangian,
(4) reduction of terms by use of various relations and (5) performing
the non--relativistic limit and working out the $1/m$ corrections by
use of the path integral~\cite{BKKM}. I will now outline these steps,
for more details I refer to the thesis of Steininger~\cite{svenphd}.
\begin{enumerate}
\item[Step 1:] We start with the fully relativistic theory, i.e. any
covariant derivative $D_\mu$ counts as order $p^0$. Formally, one
could thus construct terms of the type $\bar \Psi \ldots
D^{38}\Psi$. To avoid this, consider the covariant derivative only as 
a building block when it acts on operators sandwiched between the
spinors (the only exception to this rule is the lowest order
Lagrangian). The pertinent terms involving $D_\mu$ acting on the
nucleon fields, which are all of chiral dimension zero (or one), are given
together with the elements of the Clifford algebra in step~2 since
they are used to saturate the free Lorentz indices of the building
blocks. In this way, one avoids from the beginning all terms with an
arbitrary large number of $D_\mu$'s, which are formally allowed. With
that in mind, the possible building blocks at orders $p$, $p^2$ and
$p^3$ are (also given are the respective charge conjugation, $C$, and
parity, $P$, assignments):
\begin{eqnarray}
\begin{array}{|c|c|c|c|c|c|c|}
\hline
\mbox{Operator} & \quad C \quad & \quad P \quad & \!\!\!\!\!\!\!& 
\mbox{Operator} & \quad C \quad & \quad P \quad \\[0.05em] \hline 
\hline
& & & & & & \\[-1.4em]
u_\mu    & +  & - & &
[D_\mu, <\chi_+>] & + & +\\[0.05em] \hline 
& & & & & & \\[-1.4em]
& {\cal O}(p^2) & & &
<\hat{\chi}_- u_\mu> & + & + \\[0.05em] \hline
& & & & & & \\[-1.4em]
\hat{\chi}_+ & + & + & &
<\chi_-> u_\mu & + & + \\[0.05em] \hline
& & & & & & \\[-1.4em]
<\chi_+> & + & + & &
[\chi_-,u_\mu] & - & + \\[0.05em] \hline
& & & & & & \\[-1.4em]
\hat{\chi}_- & + & - & &
[D_\mu, \hat{\chi}_-] & + & - \\[0.05em] \hline
& & & & & & \\[-1.4em]
<\chi_-> & + & - & &
[D_\mu, <\chi_->] & + & - \\[0.05em] \hline
& & & & & & \\[-1.4em]
<u_\mu u_\nu > & + & + & &
<u_\mu u_\nu> u_\lambda & + & - \\[0.05em] \hline
& & & & & & \\[-1.4em]
[u_\mu,u_\nu] & - & + & &
<u_\mu [u_\nu,u_\lambda]> & - & - \\[0.05em] \hline 
& & & & & & \\[-1.4em]
[D_\mu,u_\nu] & + & - & &
<[D_\mu,u_\nu] u_\lambda> & + & + \\[0.05em] \hline
& & & & & & \\[-1.4em]
\hat{F}^+_{\mu\nu} & - & + & &
[[D_\mu,u_\nu],u_\lambda] & - & + \\[0.05em] \hline
& & & & & & \\[-1.4em]
<F^+_{\mu\nu}> & - & + & &
[D_\mu,[D_\nu,u_\lambda]] & + & - \\[0.05em] \hline
& & & & & & \\[-1.4em]
& {\cal O}(p^3) & & &
<\hat{F}^+_{\mu\nu} u_\lambda> & - & - \\[0.05em] \hline
& & & & & & \\[-1.4em]
<\hat{\chi}_+ u_\mu> & + & - & &
<F^+_{\mu\nu}> u_\lambda & - & - \\[0.05em] \hline
& & & & & & \\[-1.4em]
<\chi_+> u_\mu & + & - & &
[\hat{F}^+_{\mu\nu},u_\lambda] & + & - \\[0.05em] \hline
& & & & & & \\[-1.4em]
[\chi_+,u_\mu] & - & - & &
[D_\lambda,\hat{F}^+_{\mu\nu}] & - & + \\[0.05em] \hline
& & & & & & \\[-1.4em]
[D_\mu, \hat{\chi}_+] & + & + & &
[D_\lambda,<F^+_{\mu\nu}>] & - & + \\[0.05em] \hline
\end{array}
\end{eqnarray}
\noindent Here, the definition $\hat A = A -\langle A \rangle /2$
is used. Of course, hermiticity has to be assured by appropriate
factors of $i$ and combinations of terms.
Note that the terms of orders $p$ and $p^2$ are also
needed since they can come in with appropriate powers of $1/m$.

\item[Step 2:] Now we have to construct the elements of the Clifford
algebra and all other operators of dimension zero/one to contract the 
Lorentz indices of the building blocks. First, I give all operators
constructed from $\gamma$ matrices, the metric tensor and the totally
antisymmetric tensor in $d=4$ according to the number of free
indices, called $N_I$:
\beqa\label{CNi}
N_0&:& \quad 1 \,\, ,\,\, \gamma_5 \,\, ; \quad N_1 : \quad
\gamma_\mu \,\, , \, \, \gamma_\mu \gamma_5 \,\, ; \quad N_2 : \quad
g_{\mu \nu} \,\, , \,\, g_{\mu \nu} \gamma_5  \,\, , \,\, 
\sigma_{\mu \nu} \,\, , \,\, \sigma_{\mu \nu} \gamma_5  \,\, ;
\nonumber \\
N_3&:& \quad  g_{\mu \nu} \gamma_\lambda \,\, , \,\, 
 g_{\mu \nu} \gamma_\lambda \gamma_5 \,\, , \,\, 
\epsilon_{\mu\nu\lambda\alpha} \gamma^\alpha \,\, , \,\,
\epsilon_{\mu\nu\lambda\alpha} \gamma^\alpha \gamma_5 \,\, .
\eeqa
Similarly, the terms involving the covariant derivative read
\beq\label{DNi}
N_1  : \,\,\,
D_\mu \,\, ;   \quad N_2 : \,\,\,
\{D_{\mu} , D_{\nu} \} \,\, ;
\quad N_3 : \,\,\,
D_{\mu} D_{\nu} D_\lambda \, {\rm + permutations} \,\,\, .
\eeq
It is important to note that $\gamma_5$ and $g_{\mu \nu} \gamma_5$
have chiral dimension one because they only connect the large to the
small components and thus appear first at order $1/m$ (as discussed
before).
Terms with more $D_\mu$'s can always be reduced to the terms listed
in eqs.(\ref{CNi},\ref{DNi}) or only contribute to higher orders by
use of the baryon eom,
\beq\label{Neom}
{\barr D} \Psi =\biggl( m  + {g_A \over 2} \gamma_5 \barr{u} \biggr) \,
\Psi + {\cal O}(p^2) \,\, .
\eeq
Let me give one example. Be $O$ some operator of dimension three
constructed from the above building blocks and properly contracted
to be a scalar with $PC = ++$ contributing as $\bar{\Psi} \,  O \,
\Psi$ to the effective Lagrangian. So what happens to a term like
$\bar{\Psi} \,  O \, D^2 \, \Psi +{\rm h.c.}$ 
which is also of dimension three
and commensurate with all symmetries? It contributes only to higher
orders as can be seen from the following chain of manipulations:
\beqa\label{Dex}
\bar{\Psi} \,  O \, D^2 \, \Psi &=& \bar{\Psi} \,  O \, 
g_{\mu\nu} D^\mu D^\nu \, \Psi \nonumber \\
&=&\bar{\Psi} \,  O \, \barr{D} \barr{D} \, \Psi -
i \, \bar{\Psi} \,  O \, \sigma^{\mu\nu} D_\mu D_\nu \, \Psi
  \nonumber \\
&=&\bar{\Psi} \,  O \, \barr{D} \barr{D} \, \Psi -
{i \over 2} \,\bar{\Psi} \,  O \, \sigma^{\mu\nu} [D_\mu , D_\nu] \,
  \Psi \,\, . 
\eeqa
The first term on the r.h.s. of eq.(\ref{Dex}) amounts to a repeated
use of the nucleon eom, i.e. to three terms with coefficents $m^2$,
$m \,g_A$ and $g_A^2$, in order. While the first of these can 
simply be absorbed
in the LEC accompanying the original term $\bar{\Psi} O \Psi$, 
the other two start to
contribute at order four and five, respectively. Similarly, the second
term on the r.h.s. of  eq.(\ref{Dex}) leads to a dimension five operator
by use of eq.(\ref{Dcomm}).

\item[Step 3:] It is now straightforward to combine the building
blocks with the operators enumerated in step~2 to construct the
effective Lagrangian. One just has to multiply the various operators
so that the result has $J^{PC} = 0^{++}$. This gives
\beqa \label{LpiN3r}
{\cal L}_{\pi N}^{(3)} &=& \sum_{i=1}^{51} \, \bar{\Psi} \, O_i^{(3)}
\, \Psi \\
&=& \bar{\Psi} \, \gamma^\mu \, \gamma_5 \, \langle \chi_+ \, u_\mu
\rangle \, \Psi +  \bar{\Psi} \, \gamma^\mu \, \gamma_5 \, \langle
u^2 \rangle \, u_\mu \, \Psi + \bar\Psi \, \gamma_5 \, \gamma_\mu \,
[D_\mu , \chi_-] \, \Psi
\ldots \,\,\, , \nonumber
\eeqa
where the $O_i^{(3)}$ are monomials in the fields of chiral dimension three.
The full list of terms is given in \cite{svenphd} and I have 
made explicit only three terms for the later discussion. At this stage,
the Lagrangian is over--complete. All terms are allowed by the
symmetries, but they are not all independent. While one could work
with such an over--complete set of terms, it is more economical to
reduce them to the minimal number of independent ones.

\item[Step 4:] Let us now show how one can reduce the number of terms
in ${\cal L}_{\pi N}^{(3)}$. To be specific, consider the third
term on the r.h.s. of eq.(\ref{LpiN3r}). Performing partial
integrations, we can reshuffle the covariant derivative to the 
extreme left and and right,
\beq
\bar\Psi \, \gamma_5 \, \gamma_\mu \, [D_\mu , \chi_-] \, \Psi =
- \bar\Psi \, \stackrel{\leftarrow}{D}_\mu \, \gamma_5 \, \gamma_\mu
\, \chi_- \, \Psi - \bar\Psi \,  \gamma_5 \, \gamma_\mu
\, \chi_- \, D_\mu \, \Psi \,\,\, .
\eeq
Using now the eom, eq.(\ref{Neom}), and the one for $\bar{\Psi}$, this
can be cast into the form
\beq
\bar\Psi \, \gamma_5 \, \gamma_\mu \, [D_\mu , \chi_-] \, \Psi =
- 2m \, \bar\Psi \, \gamma_5 \,
\, \chi_- \, \Psi - g_A \, \bar\Psi \,  [ \barr{u} ,
\chi_- ]  \, \Psi + {\cal O}(p^4) \,\,\, 
\eeq
and these are terms which already exist at dimension three, i.e.
the term under consideration can be absorbed in the structure
of these two terms. By similar methods, one can reduce the 51 terms down
to 24 independent ones.

\item[Step 5:] As a final step, we now use the rules derived before,
see eq.(\ref{spinga}), to go to the non--relativistic limit. 
For example, the first two terms given in eq.(\ref{LpiN3r}) take
the form
\beq
\bar{H} \, S^\mu \langle \chi_+ \, u_\mu \rangle \, H +
\bar{H} \, S \cdot u \, \langle u^2 \rangle \, H \,\,\, .
\eeq
In a similar fashion, all other terms can be reduced and the $1/m$
corrections can be worked out along the lines spelled out in
appendix~A of ref.\cite{BKKM}. Let me discuss as an example the
part of the third order action which has the form~\cite{BKKM}
\beq
S_{\pi N}' = -{1 \over (2m)^2} \, \int d^4x \, \bar{H} \,( \gamma_0
\, {\cal B}^{(1)\dagger} \, \gamma_0 ) (i v \cdot D + g_A S \cdot u
) \, {\cal B}^{(1)} \, H \,\,\, ,
\eeq
which translates into the following piece of the
third order effective Lagrangian,
\beq
{\cal L}_{\pi N}^{(3)'}  = -{1 \over 4m^2} \bar{H} \,\biggl[ (i
\barr{D}^\perp + {1\over 2} g_A v \cdot u \gamma_5)(i v \cdot D + g_A
S\cdot u)( i \barr{D}^\perp - {1\over 2} g_A v \cdot u \gamma_5)
\biggr] H \,\, .
\eeq
One has now to express the various products of derivatives, $\gamma$
matrices and so on in terms of non--relativistic four--vectors like
e.g. for the product of the first terms in the three square brackets,
\beq
\bar{H}  \barr{D}^\perp  v\cdot D  \barr{D}^\perp  H
= \bar{H} \bigl[ (v \cdot D)^3 + D_\mu  v\cdot D  D^\mu - 
2[S_\mu , S_\nu]\, D^\mu  v\cdot D  D^\nu \bigr ] H \,\, ,
\eeq
so that
\beq
{\cal L}_{\pi N}^{(3)'} = -{i \over 4m^2} \bar{H} \,\biggl[ (v \cdot D)^3
- D_\mu v\cdot D D^\mu + 2[S_\mu , S_\nu] D^\mu  v\cdot D  D^\nu 
+ \ldots \biggr] H \,\, .
\eeq
With 8 other relations of this type one can fill in the
ellipsis. Similarly, the other pieces contributing at this order
can be worked out. Putting pieces togther, the complete third order 
Lagrangian takes the form
\beq \label{lpin3f}
{\cal L}_{\pi N}^{(3)} = {\cal L}_{\pi N}^{(3),\, {\rm fixed}} +
\sum_{i=1}^{24} b_i \, \bar{H} \tilde{O}_i H \,\,\, ,
\eeq
where the first term in Eq.(\ref{lpin3f}) subsumes all terms with
fixed coefficients of lower chiral dimension, i.e. the ones of the
types $g_A/m^2, 1/m^2$ and $c_i / m$, respectively. The $b_i$ are the third
order LECs. The complete Lagrangian in this basis can be found in
\cite{svenphd}\cite{mmsr} and is already given in a 
different basis in \cite{eckmoj}.

\end{enumerate}

\medskip 
\noindent
For ${\cal L}_{\pi N}^{(4)}$, no systematic study 
enumerating {\it all} terms exists so far but some
dimension four operators have been used and their corresponding  LECs
fixed \cite{bkms} \cite{bkmpi0}. Furthermore, the renormalization
of the generating functional to order $p^4$ has been worked out by
M\"uller~\cite{guidophd} and will be available soon~\cite{mmsr}.

\vfill\eject

%%%%%%%%%%%%%%%%%%%%%%%%%%%%%%%%%%%%%%%%%%%%%%%%%%%%%%%%%%%%%%%%%%%%%%
\section{Isospin symmetry and virtual photons}
\label{sec:vir}
\setcounter{equation}{0}

Up to now, I have mostly treated pure QCD in the isospin limit. We
know, however, that there are essentially two sources of isospin
symmetry violation (ISV). First, the quark mass difference $m_d -m_u$
leads to {\it strong} ISV. Second, switching on the {\it electromagnetic}
(em) interaction, charged particles are surrounded by a photon cloud
making them heavier than their neutral partners. An extreme case is
the pion, where the strong ISV is suppressed (see below) and the
photon cloud is almost entirely responsible for the charged
to neutral mass difference. Matters are different for the nucleon.
Here, pure em would suggest the proton to be heavier than the
neutron by 0.8~MeV - at variance with the data. 
However, the quark mass (strong) contribution can be estimated to be
$(m_n-m_p)^{\rm str}= 2.1$~MeV. Combining these two numbers, one arrives at the
empirical value of $m_n - m_p = 1.3$~MeV.  

\smallskip
\noindent Consider first the strong interactions. The symmetry
breaking part of the QCD Hamiltonian, i.e. the quark mass term,
 can be decomposed into an isoscalar and an isovector term
\beq
{\cal H}_{\rm QCD}^{\rm sb} = m_u \bar u u + m_d \bar d d =
{1\over 2} (m_u + m_d) ( \bar u u +  \bar d d ) +
{1\over 2} (m_u - m_d) ( \bar u u -  \bar d d ) \,\,\, .
\eeq
The quark mass ratios can easily be deduced from the ratios of the
(unmixed) Goldstone bosons, in particular, $m_d / m_u = 1.8\pm
 0.2$~\cite{leutw}. Therefore, $(m_d-m_u)/(m_d+m_u) \simeq 0.3$ and
one could expect large isospin violating effects. However, the light
quark masses are only about $5...10$~MeV (at a renormalization
scale of 1~GeV) and the relevant scale to compare
to is of the order of the proton mass. This effectively suppresses the
effect of the sizeable light quark  mass difference in most cases,
as will be discussed below.
We notice that in the corresponding meson EFT, Eq.(\ref{Lpipi}), 
the isoscalar term appears at leading order while the isovector one 
is suppressed. This is essentially the reason for the tiny quark mass 
contribution to the pion mass splitting. On the other hand, in the
pion-nucleon system, no symmetry breaking appears at lowest order
but at next--to--leading order, the isoscalar and the isovector terms
contribute. These are exactly the terms $\sim c_1$ and $\sim c_5$ in
Eq.(\ref{lp2}). In his seminal paper in 1977, Weinberg pointed out
that reactions involving nucleons and {\it neutral} pions might lead
to gross violations of isospin symmetry~\cite{weinmass} since the
leading terms from the dimension one Lagrangian are suppressed. In particular,
he argued that the mass difference of the up and down quarks can
produce a 30\% effect in the difference of the $\pi^0 p$ and $\pi^0 n$
S--wave scattering lengths while these would be equal in case of
isospin conservation. This was later reformulated in more modern
terminology~\cite{weinmit}.
To arrive at the abovementioned result, Weinberg considered
Born terms and the dimension two symmetry breakers.
However, as shown in ref.~\cite{bkmpin},
at this order there are other isospin--conserving terms which
make a precise prediction for the individual $\pi^0 p$ or $\pi^0 n$
scattering length very difficult. Furthermore, there is no way of
directly measuring these processes. On the other hand, there exists a huge
body of data for elastic charged pion--nucleon scattering ($\pi^\pm p
\to \pi^\pm p$) and charge exchange reactions ($\pi^- p \to \pi^0
n$).  In the framework of some models
it has been claimed that the presently available pion--nucleon data
basis exhibits isospin violation of the order of a few 
percent~\cite{gibbs}\cite{mats}. What is, however, uncertain is to
what extent the methods used to separate the
electromagnetic from the strong ISV effects match. To really pin down isospin
breaking due to the light quark mass difference, one needs a machinery
that allows to {\it simultaneously} treat the electromagnetic and the
strong contributions. Here, CHPT comes into the game since one can
extend the effective Lagrangian to include virtual photons as I will show
now.
 
\medskip

\noindent  Consider now the photons as dynamical degrees of freedom.
To do this in a systematic fashion, one has to extend the power
counting. A very natural way to do this is to
assign to the electric charge a chiral dimension one, 
based on the observation that
\beq
{e^2\over 4\pi} \sim {M_\pi^2 \over (4\pi F_\pi)^2} \sim {1 \over 100} \quad .
\eeq 
This is also a matter of consistency. While the neutral pion mass
is unaffected by the virtual photons, the charged pions acquire
a mass shift of order $e^2$ from the photon cloud. If one would
assign the electric charge a chiral dimension zero, then the
power counting would be messed up. This is discussed in more
detail in Gasser's lectures~\cite{juergd}.
The extension of the meson Lagrangian is standard, I only give the
result here and refer to refs.\cite{reso}\cite{ru}\cite{nr} for
all the details. Also, I  consider only the two flavor case,
\beq \label{L2meson}
{\cal L}^{(2)}_{\pi\pi} = - {1\over 4}F_{\mu\nu}F^{\mu\nu} - {\lambda \over 2} 
(\partial_\mu A^\mu )^2 + {F_\pi^2\over 4} \langle \nabla_\mu U \nabla^\mu
 U^\dagger + \chi U^\dagger + \chi^\dagger U \rangle + C \langle Q U
 Q U^\dagger \rangle \,\, , 
\eeq
with $F_{\mu\nu} = \partial_\mu A_\nu - \partial_\nu A_\mu$ the photon field
strength tensor and $\lambda$ the gauge--fixing parameter (from here on,
I use the Lorentz gauge $\lambda = 1$). Also, 
\beq
\nabla_\mu U = \partial_\mu U - i(v_\mu+a_\mu+QA_\mu)U
 +iU(v_\mu-a_\mu+QA_\mu) \,\, , 
\eeq
is the generalized pion covariant derivative containing the external
vector ($v_\mu$) and axial--vector ($a_\mu$)  sources.
It is important to stress that in
refs.~\cite{reso}\cite{ru}\cite{nr}, $Q$ denotes the {\it quark} charge
matrix. To make use of the {\it nucleon} charge matrix commonly used 
in the pion--nucleon EFT, one can
perform a transformation of the type $Q \to Q + \alpha \, e\, \bf 1$, with
$\alpha$ a real parameter. One observes that $d \langle Q U Q U^\dagger \rangle
/ d\alpha \sim e^2 \, \bf 1$, i.e. to this order the difference between
the two charge matrices can completely be absorbed in an unobservable
constant term. To use the higher order terms constructed
in~\cite{mms}, one would have to rewrite them in terms of the nucleon 
charge matrix. In what follows, I will not need these terms.
It is advantageous to work in the $\sigma$--model gauge for the pions, 
Eq.(\ref{siggauge}). In that case, the last term in
${\cal L}^{(2)}_{\pi\pi}$ leads only to a term quadratic in pion
fields. Consequently, the LEC $C$ can be
calculated from the neutral to charged pion mass difference since this
term leads to $(\delta M^2_\pi)_{\rm em} = 2e^2C/F_\pi^2$. This gives
$C = 5.9\cdot 10^{-5}\,$GeV$^{4}$. Extending this unique lowest order
term to SU(3), one can easily derive Dashen's theorem.
To introduce virtual photons in the effective pion--nucleon field
theory, consider  the nucleon charge matrix $Q = e \,{\rm diag}(1,0)$.
Note that different to what was done before, the explicit factor of
$e$ is subsumed in $Q$ so as to organize the power counting along the
lines discussed before. For the construction of chiral invariant 
operators, let me introduce the matrices
\beq
Q_{\pm} = \frac{1}{2} \, (u \, Q \,u^\dagger \pm u^\dagger \, Q \, u) \,\, ,
\quad
\hat{Q}_{\pm} = Q_\pm - {1 \over 2} \langle Q_\pm \rangle \,\, .
\eeq
By construction, the
$\hat{Q}_{\pm}$ are traceless. Under chiral SU(2)$_L\times$SU(2)$_R$
symmetry, the $Q_\pm$ transform as 
\beq
Q_\pm  \to K \, Q_\pm  \, K^\dagger \quad .
\eeq 
Furthermore, under parity ($P$) and charge conjugation ($C$)
transformations, one finds
\beq
P \, Q_\pm \, P^{-1} = \pm \, Q_\pm \,\,\, , \quad 
C \, Q_\pm \, C^{-1} = \pm \, Q_\pm^T \,\,\, , 
\eeq
where $Q^T$ is the transposed of the matrix $Q$. For physical
processes, only quadratic combinations of the charge matrix $Q$
(or, equivalently, of the matrices $Q_\pm$) can appear. The following
relations are of use,
\beq
\langle Q_- \rangle = \langle Q_- \, Q_+ \rangle = 0 \,\, , \quad
\langle [i D_\mu , Q_\pm ] \rangle = 0 \,\,\, ,
\eeq
together with the SU(2) matrix relation $\{A,B\} = \langle AB \rangle 
+ A \langle B \rangle  + B  \langle A \rangle -  \langle A \rangle
 \langle B \rangle$. It is now
straightforward to implement the (virtual) photons given in terms
of the gauge field $A_\mu$ in the effective pion--nucleon Lagrangian.
Starting from the relativistic theory and decomposing the spinor
fields into light (denoted $N$) and heavy components (velocity
eigenstates), one can proceed as discussed before.
In particular, to lowest order (chiral dimension one)
\beq
{\cal L}_{\pi N}^{(1)} = \bar{N} \,\biggl( i v \cdot \tilde{D} +
g_A \, S \cdot \tilde{u} \, \biggr) \, N\,\,\, , 
\eeq
with 
\beq \label{covder}
\tilde{D}_\mu = D_\mu -i \, Q_+ \, A_\mu \,\,\, , \quad 
\tilde{u}_\mu = u_\mu - 2 \, Q_- \, A_\mu \,\,\, .
\eeq
At next order (chiral dimension two), we have 
\beq 
{\cal L}_{\pi N, {\rm em}}^{(2)} = \bar{N} \, F_\pi^2 \, \biggl\{ {f_1} \,
\langle Q^2_+ - Q^2_- \rangle + {f_2} \, \hat{Q}_+ \langle Q_+ \rangle
+ {f_3} \, \langle Q^2_+ + Q^2_- \rangle +
{f_4} \ \langle Q_+\rangle^2 \, \biggr\} \, N \,\,\, .
\eeq
For the electromagnetic terms, we have written down the
minimal number allowed by all symmetries. Note that the last two terms
in ${\cal L}_{\pi N, {\rm em}}^{(2)}$ are proportional to $e^2 \bar{N}
N$. This means that they
only contribute to the electromagnetic nucleon mass in the chiral
limit and are thus not directly observable. However, this implies that
in the chiral two--flavor limit ($m_u =m_d=0$, $m_s$ fixed), the
proton is heavier than the neutron since it acquires an
electromagnetic mass shift. Only in pure QCD ($e^2=0$), this 
chiral limit mass is the same for the two particles.
Since at present only calculations to order
$q^3$ have been performed, one can always use the physical nucleon
mass, denoted $m$, and do not need to bother about its precise value
in the chiral limit. The numerical values of the electromagnetic LECs
$f_1$ and $f_2$ will be discussed below. The normalization factor of
$F_\pi^2$ in the electromagnetic pion--nucleon Lagrangian is
introduced so that the $f_i$ have the same dimension as the strong
LECs $c_i$. 

\medskip

\renewcommand{\arraystretch}{1.2}
\begin{table}[hbt]
\begin{center}
\begin{tabular}{|c|c|c|c|c|}
   \hline
    Operator    & $P$   &  $C$  & $\Gamma_\mu$  &  HBCHPT     \\
    \hline
    $\langle Q_+^2\rangle u_\mu $  &  $-$  & $+$ & $\gamma_5 \gamma_\mu$ & 
    $\langle Q_+^2\rangle S \cdot u $ \\ 
    $[Q^2 , u_\mu] $  &  $-$  & $-$ & no  & --- \\ 
    $\langle[Q_+ , Q_-]u_\mu \rangle $  &  $+$  & $+$ & no  & --- \\ 
    $[Q_+ , [Q_- ,u_\mu ]] $  &  $+$  & $-$ & $\gamma_\mu$  
    & $[Q_+ , [Q_- ,v \cdot u ]]    $ \\ 
    $\langle Q_+^2\rangle i D_\mu$ &  $+$  & $-$ & $ \gamma_\mu$ &
    $\langle Q_+^2\rangle i v \cdot D$ \\ 
    $\langle Q_+ \rangle [iD_\mu,Q_+] $  &  $+$  & $+$ & no  & --- \\ 
    $[Q_+ ,[iD_\mu,Q_+]] $  &  $+$  & $-$ & $\gamma_\mu$  & 
    $[Q_+ ,[i v \cdot D ,Q_+]] $                       \\ 
    $[Q_+ , Q_-] iD_\mu $  &  $-$  & $-$ & no  & --- \\ 
    $\langle Q_+ \rangle [ Q_-,iD_\mu] $  &  $-$  & $-$ & no  & --- \\ 
    $[Q_+ , [ iD_\mu , Q_-]] $  &  $-$  & $+$ & $\gamma_5 \gamma_\mu$
    & $[Q_+ , [ i S \cdot D , Q_-]] $  \\ 
    \hline
\end{tabular}
\end{center}
\caption{Contruction of the dimension three operators including
    virtual photons. Each of these operators represents a class
    of operators with the same behaviour under parity and charge
    conjugation.}
\end{table}
\noindent To go beyond tree level, we have to construct the
terms of order $p^3$. From the building blocks at our disposal, we can
construct 17 independent terms which are compatible with all
symmetries, following ref.~\cite{krause}. The number of possible terms
is limited due to the fact that at least two charge matrices must
appear. In particular, charge
conjugation allows to sort out terms which would otherwise be allowed.
In table~2, I write down a typical set of operators, starting
from the relativistic theory and then transforming  into the
heavy baryon formalism. The Dirac matrix
$\Gamma_\mu$ has to be chosen such that a) the indices get contracted
and b) the corresponding operator has $PC=++$. In some case, denoted
by 'no', it is not possible to achieve this with either $\gamma_\mu$,
$\gamma_5$ or $\gamma_\mu \gamma_5$ (these are the only Dirac matrices
which can be used to construct a dimension three term fulfilling the
aforementioned requirements). Some of these terms are
accompanied by finite LECs whereas the others absorb the divergences
appearing at one loop with LECs that are only finite after
renormalization. The renormalization procedure to render these finite
is spelled out in detail in ref.\cite{suiso}.
The electromagnetic part of the 
dimension three  Lagrangian takes the form (after combining all finite
terms with the ones obtained after renormalization)
\beq \label{L3em}
{\cal L}^{(3)}_{\pi N,{\rm em}} = \sum_{i=1}^{17} g_i \, \bar N \,
{\cal O}_i \, N \,\, , 
\eeq
with the ${\cal O}_i$ monomials in the fields of dimension three. The
low--energy constants $g_i$ absorb the divergences in the standard
manner,
\beq \label{L3LECs}
g_i = \kappa_i \, L + g_i^r (\mu ) \, \,\,\,\, ,
\eeq
with $\mu$ the scale of dimensional regularization.
% and $d$ the number of space--time dimensions. 
\renewcommand{\arraystretch}{1.0}
\begin{table}[t]\begin{center}
\begin{tabular}{|r|c|c|} \hline
i  &  ${\cal O}_i$  &          $\kappa_i$ \\ \hline
1  & $ \langle Q_+ \, S \cdot u\rangle \, Q_+  $ & $ 4\,Z\,g_A (1-g_A^2) $
 \\
2  & $ \langle Q_- \, S \cdot u\rangle \, Q_-  $ & $ 2\,g_A (1-g_A^2)
(1- 2Z) $ \\
3  & $ \langle Q_+ \, S \cdot u\rangle \, \langle Q_+\rangle   $ & $ -2\,g
_A
(1+Z-Z\,g_A^2) $ \\
4  & $ \langle Q_+^2 - Q_-^2\rangle \, S \cdot u  $ & $ g_A (8Z-g_A^2)/2 $
 \\
5  & $ \langle Q_+^2 + Q_-^2\rangle \, S \cdot u  $ & $ -g_A (4-g_A^2)/2$ 
\\
6  & $ \langle Q_+\rangle ^2 \, S \cdot u  $ & $ -2 Z\,g_A $ \\
7  & $ \langle Q_- \, v \cdot u\rangle \, \langle Q_+\rangle   $ 
   & $ 1- 3\,g_A^2 $ \\
8  & $ \langle Q_- \, v \cdot u\rangle \, Q_+  $ & $ -2( 1- 3\,g_A^2) $ \\
9  & $ \langle Q_+ \, v \cdot u\rangle \, Q_-  $ & $  -4 $ \\
10 & $ \langle Q_+\rangle \, Q_+ \, i v \cdot D + \mbox{h.c.}  $ & $ -2 $ 
\\
11 & $ \langle Q_+^2 - Q_-^2\rangle \, i v \cdot D  + \mbox{h.c.}  $ & $
-1/2-3\,g_A^2/4 (1+8Z) $ \\
12 & $ \langle Q_+^2 + Q_-^2\rangle \, i v \cdot D  + \mbox{h.c.}  $ & $ -
1/2 +
3\,g_A^2 /4 $ \\
13 & $ \langle Q_+\rangle ^2 \, i v \cdot D  + \mbox{h.c.} $ & $ 1+3Z\,g_A
^2 $ \\
14 & $ [Q_+,i v \cdot c_+]  $ & $ -2 $ \\
15 & $ [Q_-,i v \cdot c_-]  $ & $ -(1-9\,g_A^2)/2 $ \\
16 & $ [Q_+,i S \cdot c_-]  $ & $ 0 $ \\
17 & $ [Q_-,i S \cdot c_+]  $ & $ 0 $ \\
\hline
\end{tabular}
\caption{Operators of dimension three and their $\beta$--functions.
Here, $Z = C/F_\pi^4$.}
\end{center}
\end{table}
\noindent The explicit expressions for the operators
${\cal O}_i$ and their $\beta$--functions $\kappa_i$ are collected in 
table~3. The $g_i^r (\mu)$ are the renormalized, finite and
scale--dependent low--energy constants. These can  be fixed by data or
have to be estimated with the help of some model. They obey the
standard renormalization group equation,
\beq
g_i^r (\mu_1) = g_i^r (\mu_2) + \kappa_i \, \log {\mu_2 \over \mu_1} 
\quad .
\eeq
 %To arrive at the terms in the table, I have used the relation
%\beqa 
%[D_\mu , Q_\pm] &=& -{i \over 2} [u_\mu , Q_\mp ] + c_\mu^\pm \,\,\, 
%\nonumber \\
%c_\mu^\pm &=& {1\over 2} \biggl\{ u (\partial_\mu Q - i [v_\mu-a_\mu ,
%Q]) u^\dagger \pm u^\dagger (\partial_\mu Q - i [v_\mu +a_\mu ,Q])u 
%\biggr\} \,\,\,\, .
%\eeqa
%Furthermore, one could use the relation
%\begin{equation}
%[\nabla^\mu,u_\mu] = \frac{i}{2} \, \chi_- - \frac{i}{4} \,
%\langle\chi_- \rangle + i\frac{4C}{F^2}\, [Q_+,Q_-]
%+ {\cal O}(q^4) \,\, ,
%\end{equation}
%to rewrite some of the terms tabulated. In fact, this has been done by
%Ecker~\cite{ecker} for the case without virtual photons ($C=0$).
%We prefer not to do this and therefore Ecker's operator ${\cal
%  O}_8^{\rm str} =[\chi_-,v \cdot u]$ in our basis 
%reads ${\cal O}_8^{\rm str} 
%= -2 [ [i\nabla^\mu,u_\mu], v \cdot u]$. 
A few remarks concerning the operators given in table~3 are in order.
First, ${\cal O}_5$ and ${\cal O}_6$ only lead to an electromagnetic
renormalization of $g_A$ and their effects can thus completely be absorbed
in the physical value of the axial--vector coupling constant. The operators
${\cal O}_{14,\ldots,17}$ are only of relevance for processes with external
axial--vector fields (the $c_\pm$ are defined in ref.\cite{suiso})
 and ${\cal O}_{10,\ldots,13}$ can be eliminated by
use of the nucleon equations of motion, i.e. they are not relevant for
reactions involving on--shell nucleons.
The general effective pion--nucleon Lagrangian with virtual
photons constructed here allows now to systematically investigate the
influence of isospin--breaking due to the quark mass difference $m_u -
m_d$ and the dual effects from electromagnetism. Of particular
importance are the processes $\pi N \to \pi N$, $\gamma N \to \pi N$
and $\pi N \to \pi \pi N$ since a large body of precise low--energy
data exists which can be analysed within the framework outlined. 

\medskip

\noindent Consider now the neutron-proton mass difference
as an instructive example. It is given
by a strong insertion $\sim c_5$ and an electromagnetic insertion
$\sim f_2$,
\beq \label{delnp}
m_n - m_p = (m_n - m_p)_{\rm str} + (m_n - m_p)_{\rm em} = 
4 \,c_5\, B \, (m_u - m_d) + 2 \, e^2 \, F_\pi^2 \, f_2  
+{\cal O}(p^4) \,\,\, .
\eeq
Note that one--loop corrections only start at  order $p^4$.
This can be traced back to the fact that the
photonic self--energy diagram of the proton (on--shell) 
at order $p^3$ vanishes
since it is proportional to $\int d^dk \, [k^2 \, v\cdot k]^{-1}$.
Such an integral vanishes in dimensional regularization.  At chiral
dimension three, the  electromagnetic LEC $f_2$ can therefore be
fixed from the electromagnetic proton mass shift, $ (m_n - m_p)_{\rm
em} = -(0.7\pm 0.3)\,$MeV, i.e. $f_2 = -(0.45\pm 0.19)\,$GeV$^{-1}$.
The strong contribution has been used in~\cite{bkmlec} to fix the LEC
$c_5 = -0.09 \pm 0.01\,$GeV$^{-1}$. Note that it is known that
one--loop graphs with an insertion $\sim m_d-m_u$ on the internal 
nucleon line, which in our counting appear at fourth order, can
contribute sizeably to the strong neutron-proton mass
difference~\cite{gl82}. This underlines the need for a complete ${\cal
  O}(p^4)$ calculation.

%%%%%%%%%%%%%%%%%%%%%%%%%%%%%%%%%%%%%%%%%%%%%%%%%%%%%%%%%%%%%%%%%%%%%%%%%%
\section{The meaning of low--energy theorems (LETs)}
\label{sec:LET}
\setcounter{equation}{0}

In this section, I will briefly discuss the meaning of the so--called
low--energy theorems. More details are given in ref.\cite{gerulf}. 
Let us first consider a well--known example of a LET
 involving the electromagnetic current. Consider the
scattering of very soft photons on the proton, i.e., the Compton scattering
process $\gamma (k_1) + p(p_1) \to \gamma (k_2) + p(p_2)$ 
and denote by $\ve \, (\ve ')$
the polarization vector of the incoming (outgoing) photon. The transition
matrix element $T$ (normalized to $d\sigma / d\Omega = |T|^2$) can be 
expanded in a Taylor series in the small parameter $\delta =
|\vec{k_1}|/m$. 
In the forward  direction and in a gauge where
the polarization vectors have only space components, $T$ takes the form
\begin{equation}
T = c_0 \, \vec{\ve}\, ' \cdot \vec{\ve} + i \, c_1 \, \delta 
\, \vec{\sigma} \cdot (
\vec{\ve}\, ' \times \vec{\ve} \, ) + {\cal O}(\delta^2) ~ .
\label{Comp}
\end{equation}
The parameter $\delta$ can be made arbitrarily small in the laboratory so
that the first two terms in the Taylor expansion (\ref{Comp}) dominate. 
To be precise, the first one proportional to $c_0$ gives the low--energy 
limit for the spin--averaged Compton amplitude, while the second ($\sim c_1$) 
is of pure spin--flip type and can directly be detected in polarized 
photon proton scattering (to my knowledge, such a test has not yet
been performed). The pertinent LETs fix the values of $c_0$ and 
$c_1$ in terms of measurable quantities \cite{low},
\begin{equation}
c_0 = - \frac{Z^2e^2}{4 \pi m} \, , \quad c_1 = 
- \frac{Z^2e^2 \kappa_p^2}{8 \pi m}
\label{c01}
\end{equation}
with $Z =1$ the  charge of the proton. To arrive at Eq.~(\ref{c01}), 
one only makes use of gauge 
invariance and the fact that the $T$--matrix can be written in terms of a
time--ordered product of two conserved vector currents sandwiched between
proton states. The derivation proceeds by showing that for small
enough photon energies the matrix element is determined by the electromagnetic
form factor of the proton at $q^2 = 0$ \cite{low}.

Similar methods can be applied to other than the electromagnetic
currents. In strong interaction physics, a special role is played by the 
axial--vector currents. The associated symmetries are spontaneously
broken giving rise to the Goldstone matrix elements
\beq
\langle 0|A^a_\mu(0)|\pi^b(p)\rangle =
i \delta^{ab} F_\pi p_\mu
\label{Gme} 
\eeq
where $a,b$ are isospin indices. In the chiral limit,
the massless pions play a similar role as the photon and many
LETs have been derived for ``soft pions".
In light of the previous discussion on Compton
scattering, the most obvious one is Weinberg's prediction for elastic 
$\pi p$ scattering \cite{wein68}. We only need the following translations~:
\begin{equation}
\langle p| T \, j_\mu^{\rm em} (x) j_\nu^{\rm em} (0)|p\rangle
 \, \, \to \, \,
\langle p| T \, A_\mu^{\pi^+} (x) A_\nu^{\pi^-} (0)|p \rangle ~, 
%\label{e8a}
\end{equation}
\begin{equation}
\partial^\mu j_\mu^{\rm em} = 0 \, \, \to \, \,
\partial^\mu A_\mu^{\pi^-}  = 0  ~.
%\label{e8b}
\end{equation}
In contrast to photons, pions are not massless in the real
world. It is therefore interesting to find out how the LETs for
soft pions are modified in the presence of non--zero pion masses
(due to non--vanishing quark masses). In the old days of current
algebra, a lot of emphasis was put on the PCAC
(Partial Conservation of the Axial--Vector Current) relation,
consistent with the Goldstone matrix element (\ref{Gme}),
\begin{equation}
\partial^\mu A^a_\mu = M_\pi^2 F_\pi \pi^a ~ \quad .
\label{PCAC}
\end{equation}
Although the precise meaning of (\ref{PCAC}) has
long been understood \cite{col}, it does not offer a systematic
method to calculate higher orders in the momentum and mass expansion
of LETs. The derivation of non--leading terms in the days of
current algebra and PCAC was more an art than a science, often
involving dangerous procedures like off--shell
extrapolations of amplitudes. In the modern language, i.e. the EFT of
the Standard Model, these higher order corrections can be calculated 
unambigously and one correspondingly defines a low--energy theorem via:

\smallskip

\begin{displaymath}
 {\rm{\bf L}(OW)} \quad {\rm {\bf E}(NERGY)} \quad{\rm  {\bf
 T}(HEOREM)} \quad {\rm OF} \quad  {\cal O}(p^n) \end{displaymath}
\beq \equiv {\rm GENERAL} \quad {\rm PREDICTION} \quad
 {\rm OF} \quad {\rm CHPT} \quad {\rm TO} \quad {\cal O}(p^n)  \quad . \eeq
\noindent 
By general prediction I mean a
strict consequence of the SM depending on some LECs like 
$F_\pi, m, g_A, \kappa_p, \ldots$, but without any model assumption for these
parameters. This definition contains a
precise prescription how to obtain higher--order corrections to 
leading--order LETs.

The soft--photon theorems, e.g., for Compton scattering \cite{low},
involve the limit of small photon momenta, with all other momenta
remaining fixed. Therefore, they hold to all orders in the non--photonic
momenta and masses. In the low--energy expansion of CHPT, on the
other hand, the ratios of all small momenta and pseudoscalar meson
masses are held fixed. Of course, the soft--photon theorems are also
valid in CHPT as in any gauge invariant quantum field theory. 
To understand this difference  of low--energy limits,
I will now rederive and extend the LET for spin--averaged nucleon
Compton scattering in the framework of HBCHPT \cite{BKKM}. Consider the
spin--averaged Compton amplitude in forward direction (in the Coulomb
gauge $\ve \cdot v = 0$) 
\beq
e^2 \ve^\mu \ve^\nu
\frac{1}{4} {\rm Tr} \biggl[ (1 + \gamma_\lambda v^\lambda) T_{\mu \nu} (v,k)
\biggr] = e^2 \biggl[ \ve^2 U(\omega ) + (\ve \cdot k )^2
V(\omega ) \biggr]
\eeq
with $\omega = v \cdot k$ ($k$ is the photon momentum) and
\beq
T_{\mu \nu} (v,k) = \int d^4 k \, {\rm e}^{ik \cdot x} \, \langle
 N(v)| T j^{\rm em}_\mu (x) j_\nu^{\rm em} (0) |N(v)\rangle~.
\eeq
All dynamical information is
contained in the functions $U(\omega)$ and $V (\omega )$. We
only consider $U(\omega)$ here and refer to Ref.~\cite{BKKM} for
the calculation of both $U(\omega)$ and $V(\omega)$. In the
Thomson limit, only $U(0)$ contributes to the amplitude.
In the forward direction, the only quantities with non--zero chiral
dimension are $\omega$ and $M_\pi$. In order to make this dependence
explicit, we write $U(\omega,M_\pi)$ instead of $U(\omega)$. With
$N_\gamma = 2$ external photons, the degree of homogeneity $D_L$
for a given CHPT contribution to $U(\omega,M_\pi)$ follows from 
Eq.~(\ref{DLMB})~:
\beq
D_L =2L - 1  + \sum_d (d-2)  N_d^M + \sum_d (d-1)  N_d^{MB}  \, \ge - 1 ~ .
\label{DLC}
\eeq
Therefore, the chiral expansion of $U(\omega,M_\pi)$ takes the following
general form~:
\beq
U(\omega,M_\pi) = \sum_{D_L\ge -1} \omega^{D_L} f_{D_L}(\omega/M_\pi)~.
\label{Uce}
\eeq
The following arguments illuminate the difference and the
interplay between the soft--photon limit and the low--energy expansion
of CHPT. Let us consider first the leading terms in the chiral expansion
(\ref{Uce})~:
\beq
U(\omega,M_\pi) = {1\over \omega} f_{-1}(\omega/M_\pi) +
 f_0(\omega/M_\pi) + {\cal O}(p)~.
\eeq
Eq.~(\ref{DLC}) tells us that only tree diagrams can contribute
to the first two terms. However, the relevant tree diagrams
do not contain pion lines. Consequently, the functions
$f_{-1}$, $f_0$ cannot depend on $M_\pi$ and are therefore constants.
Since the soft--photon theorem \cite{low} requires $U(0,M_\pi)$
to be finite, $f_{-1}$ must actually vanish and the chiral
expansion of $U(\omega,M_\pi)$ can be written as
\beq
U(\omega,M_\pi) = f_0 + \sum_{D_L\ge 1} \omega^{D_L} f_{D_L}(\omega/M_\pi)~.
\label{Uce2}
\eeq
But the soft--photon theorem yields additional information~:
since the Compton amplitude is independent of $M_\pi$ in the Thomson
limit and since there is no term linear in $\omega$ in the spin--averaged
amplitude, we find
\beq
\lim_{\omega \to 0}~ \omega^{n-1} f_n(\omega/M_\pi) = 0 \qquad
(n\ge 1) \label{spl}
\eeq
implying in particular that the constant $f_0$ describes the
Thomson limit~:
\beq
U(0,M_\pi) = f_0~.
\eeq
Let me now verify these results by explicit calculation.
In the Coulomb gauge, there is no direct 
photon--nucleon coupling from the lowest--order effective Lagrangian 
${\cal L}_{\pi N}^{(1)}$ since it is proportional to $\ve \cdot v$. 
Consequently, the corresponding Born diagrams  vanish so that indeed
$f_{-1}=0$. On the other hand, I had argued in section \ref{sec:lpin2} that
the heavy mass expansion of the relativistic $\pi N$ Dirac Lagrangian 
leads to a Feynman insertion of the form (from the first two terms in 
eq.(\ref{lp2})):
\beq
i \frac{e^2}{m} \frac{1}{2} (1 + \tau_3 ) \biggl[ \ve^2 -( \ve
\cdot v)^2 \biggr] = i \frac{e^2 Z^2}{m} \,  \ve^2
\eeq
producing the desired result $f_0 = Z^2 / m$, the Thomson limit.

At the next order in the chiral expansion, ${\cal O}(p^3)$ ($D_L = 1$), 
the function $f_1(\omega/M_\pi)$ is given by the
finite sum of 9 one--loop diagrams \cite{BKM} \cite{BKKM}. According
to Eq.~(\ref{spl}), $f_1$ vanishes for $\omega \to 0$. The term linear
in $\omega/M_\pi$ yields the leading contribution to the sum of the
electric and magnetic polarizabilities of the nucleon, defined by
the second--order Taylor coefficient in the expansion of $U(\omega,M_\pi)$
in $\omega$~:
\beq
f_1(\omega/M_\pi) = - {11 g_A^2 \omega\over 192 \pi F_\pi^2 M_\pi}
+ {\cal O}(\omega^2)~.
\eeq
The $1/M_\pi$ behaviour should not come as a surprise
-- in the chiral limit the  pion cloud becomes long--ranged (instead of being
Yukawa--suppressed) so that the polarizabilities explode.
This behaviour is specific to the leading contribution
of ${\cal O}(p^3)$. In fact, from the general form (\ref{Uce2}) one
immediately derives that the contribution of ${\cal O}(p^n)$
($D_L = n - 2$) to the polarizabilities is of the form $c_n M_\pi^{n-4}$
($n\ge 3$), where $c_n$ is a constant that may be zero.
One can perform a similar analysis for the amplitude 
$V(\omega)$ and for the spin--flip amplitude. We do not
discuss these amplitudes here but refer the reader to Ref.~\cite{BKKM} 
for details.

Next, let us consider the processes
$\gamma N \to \pi^0 N$ $(N=p,n)$
at threshold, i.e., for vanishing three--momentum of the pion
in the nucleon rest frame. At
threshold, only the electric dipole amplitude $E_{0+}$ survives and
the only quantity with non--zero chiral dimension is $M_\pi$.
In the usual conventions,
$E_{0+}$ has physical dimension $-1$ and it can therefore be
written as
\beq
E_{0+} = {e g_A \over F}~A\left( {M_\pi\over m},{M_\pi\over F}\right)~.
\eeq
The dimensionless amplitude $A$ will be expressed as a power
series in $M_\pi$. The various parts are characterized by the
degree of homogeneity (in $M_\pi$) $D_L$ according to the chiral
expansion. Since $N_\gamma=1$ in the present case, we obtain from
Eq.~(\ref{DLMB})
\beq
D_L = D-1 = 2L + \sum_d (d-2)  N_d^M + \sum_d (d-1)  N_d^{MB} ~ .
\eeq
For the LET of ${\cal O}(p^3)$ in question, only lowest--order mesonic vertices
($d=2$) will appear. Therefore, in this case the general formula for $D_L$ 
takes the simpler form 
\beq
D_L = 2L + \sum_d (d-1)  N_d^{MB} ~ .
\label{DLphoto}
\eeq
I will not discuss the chiral expansion of $E_{0+}$ step by step, referring
to the literature \cite{BGKM} \cite{BKM1} \cite{BKKM} for the actual 
calculation and for more details to the Comment
\cite{gerulf}. Up--to--and--including order $\mu^2$, one has to
consider contributions with $D_L = 0, 1$ and $2$. In fact, for neutral
pion photoproduction, there is no term with $D_L=0$ since the
time--honored Kroll--Ruderman contact term
\cite{KR} where both the pion and the photon emanate from the same
vertex, only exists for charged pions.
For $D_L = 2$
there is a one--loop
contribution ($L=1$) with leading--order vertices only
($N_d^{MB}=0 ~(d>1)$).  It is considerably easier to work out the
relevant diagrams in HBCHPT \cite{BKKM} than in the original derivation
\cite{BGKM} \cite{BKM1}. In fact, at threshold only the so--called triangle
diagram (and its crossed partner) survive out
of some 60 diagrams. The main reason for the enormous simplification
in HBCHPT is that one can choose a gauge without a direct $\gamma NN$
coupling of lowest order and that there is no direct coupling of
the produced $\pi^0$ to the nucleon at threshold. 
Notice that the loop contributions are finite and they are identical for proton
and neutron. They were omitted in the original version of the LET
\cite{VZ} \cite{deB} and in many later rederivations.
The full LETs of ${\cal O}(p^3)$ are given by \cite{BGKM} 
\beq
E_{0+}(\pi^0 p)=  - \dfrac{e g_A}{8\pi F_\pi}\biggl[ \biggr.
\dfrac{M_\pi}{m} - \dfrac{M_\pi^2}{2 m^2}~(3+\kappa_p)  -
\dfrac{M_\pi^2}{16 F_\pi^2} + \biggl.{\cal O}(M_\pi^3)\biggr] \eeq
\beq
E_{0+}(\pi^0 n)=  - \dfrac{e g_A}{8\pi F_\pi}\biggl[ \biggr. 
 \dfrac{M_\pi^2}{2 m^2}~\kappa_n  -
\dfrac{M_\pi^2}{16 F_\pi^2} + \biggl. {\cal O}(M_\pi^3)\biggr] \eeq
We note that the electric dipole amplitude for neutral pion production
vanishes in the chiral limit. By now, even the terms of order
$M_\pi^3$ have been worked out, see ref.\cite{bkmpi0}.
The derivation of LETs sketched above is based on a well--defined
quantum field theory where each step can be checked explicitly. Nevertheless, 
the corrected LETs have been questioned by several authors. A detailed
discussions of the various assumptions like e.g. analyticity in the
pion mass, which do not hold in QCD, can be found in
ref.\cite{gerulf}. Even better, by now the data support the
CHPT predictions, see section~11.1.

\medskip

\noindent The last example I want to treat is
the case of two--pion photoproduction. At threshold, the 
two--pion photoproduction current matrix element can be decomposed
into amplitudes as follows (to first order in $e$ and in the gauge $
\epsilon_0=0$):
\beq T \cdot \epsilon = \chi^\dagger_f \left\lbrace i \vec \sigma \cdot
(\vec \epsilon \times \vec k \,) [ M_1 \delta^{ab} +
M_2 \delta^{ab} \tau^3 + M_3 (\delta^{a3} \tau^b + \delta^{b3} \tau^a)] 
\right\rbrace \, \chi_i \, , \eeq
with $\chi_{i,f}$ two--component Pauli and isospinors. Here, I am only 
interested in the first non--vanishing contributions to $M_{2,3}$,
given by some tree diagrams. So we have $N_\gamma =1$ , $L=0$ and only
lowest order mesonic vertices ($d=2$), i.e.
\beq D_L = \sum_d (d-1) N_d^{MB} \quad . \eeq
Tree diagrams with lowest order vertices from ${\cal L}_{\pi N}^{(1)}$
all vanish due to the threshold selection rules $\epsilon \cdot v =
\epsilon \cdot q_i =0$, $S \cdot q_i = 0$ and $v \cdot (q_1 -q_2) =
0$. The first non--vanishing contribution comes from tree diagrams
with insertions from ${\cal L}_{\pi N}^{(2)}$, in particular the
expansion of $f_{\mu \nu}^+$ from eq.(\ref{lp2}) leads to a $\gamma
\pi \pi NN$ vertex proportional to $\krig{\kappa}_V$ so that
\beq M_2 = -2M_3 = \frac{e}{4 \krig{m} F^2} \left( 2 \krig{g}_A^2 - 1 -
\krig{\kappa}_V \right) + {\cal O}(p) 
= \frac{e}{4 \ m F_\pi^2} \left( 2 g_A^2 - 1 -
\kappa_V \right) + {\cal O}(p) \, \, . \eeq
The full one--loop results can be found in ref.\cite{bkm2p}. It is
amusing to note that this particular vertex has been overlooked in
other  derivations of these LETs \cite{dd} \cite{bt} simply because
not the most {\it general} effective $\pi N$ Lagrangian at order $p^2$
was used.

%%%%%%%%%%%%%%%%%%%%%%%%%%%%%%%%%%%%%%%%%%%%%%%%%%%%%%%%%%%%%%%%%%%%%%%%%%
\section{Spectral functions of electroweak form factors}
\label{sec:SF}
\setcounter{equation}{0}

As a first application, I will now take a closer look at the spectral
functions of the nucleon's electroweak form factors at low $t$ since
CHPT can be applied to extract some useful information.
The structure of the nucleon  as probed with virtual photons
is parametrized in terms of four form factors (here, $N$ is the fully
relativistic spin--1/2 field),
\beq
<N(p')\, | \, {\cal J}_\mu \,  | \, N(p)> 
= e \,  \bar{u}(p') \, \biggl\{  \gamma_\mu F_1^{N} (t)
+ \frac{i \sigma_{\mu \nu} k^\nu}{2 m} F_2^{N} (t) \biggr\} 
\,  u(p) \,, \quad N=p,n \,,
\eeq
with $t = k_\mu k^\mu = (p'-p)^2$ the invariant momentum 
transfer squared and ${\cal J}_\mu$ 
the em current related to the photon field.
In electron scattering, $t < 0$ and it is thus convenient 
to define the positive
quantity $Q^2 = -t > 0$. $F_1$ and $F_2$ are called the Pauli and the Dirac
form factor (ff), respectively, with the normalizations $F_1^p (0) =1$,
$F_1^n (0) =0$, $F_2^p (0) =\kappa_p$ and $F_2^n (0) =\kappa_n$. 
Also used are the electric and magnetic Sachs ffs,
\beq
G_E = F_1 - \tau F_2 \, , \quad G_M = F_1 + F_2 \, ,
 \quad \tau =Q^2/4m^2 \,.
\eeq
In the Breit--frame, $G_E$ and $G_M$ are nothing but the Fourier--transforms
of the charge and the magnetization distribution, respectively.
There exists already a large body of data
for the proton and also for the neutron. 
In the latter case, one has to perfrom
some model--dependent extractions to go from the deuteron
or $^3$He to the neutron. 
More accurate data are soon coming (ELSA, MAMI, TJNAF, $\dots$). 
There are also data in the time--like region from the reactions $e^+ e^-
\to p \bar{p} , n \bar{n}$ and from annihilation $p \bar{p} \to
e^+ e^-$, for $t \ge 4m_N^2$. It is
thus mandatory to have a method which allows to analyse all these data in a
mostly model--independent fashion. That's were dispersion theory comes into
play. Although 
not proven strictly (but shown to hold in all orders in perturbation
theory), one writes down an unsubtracted dispersion relation for $F(t)$ (which
is a generic symbol for any one of the four ff's),
\beq
F(t) = \frac{1}{\pi} \int_{t_0}^\infty \, dt' \, \frac{{\rm Im} \, F(t)}{t'-t}
\, \, , \eeq
with $t_0$ the two (three) pion threshold for the isovector (isoscalar) ffs.
Im~$F(t)$ is called the {\it spectral}
{\it  function}. It is advantageous to work in
the isospin basis, $F_i^{s,v} = (F_i^p \pm F_i^n)/2$, since the photon
has an isoscalar ($I=s$) and an isovector ($I=v$) component. These spectral 
functions are the natural meeting ground for theory and experiment, 
like e.g. the partial
wave amplitudes in $\pi N$ scattering. In general, the spectral functions 
can be thought  of as a superposition of vector meson poles and some 
continua, related to n-particle thresholds, like
e.g. $2\pi$, $3\pi$, $K \bar{K}$, $N\bar{N}$ and so on. 
For example, in the Vector Meson
Dominance (VMD) picture one simply retains a set of poles.  
It is important to realize that  there are some
powerful constraints which the spectral functions have to obey. 
Consider the spectral functions just above threshold.
Here, {\it unitarity} plays a central role. As pointed out by Frazer and 
Fulco  \cite{frafu}
long time ago, extended unitarity leads to a drastic enhancement 
of the isovector spectral functions on the left wing of the $\rho$
resonance. This is due to a logarithmic singularity on the second
Riemann sheet at the {\it anomalous} threshold
\beq t_c = 4M_\pi^2 - {M_\pi^4 \over m^2} = 3.98 M_\pi^2 \quad ,\eeq
very close to the normal threshold $t_0 = 4M_\pi^2$. Leaving out this
contribution from the two--pion cut leads to a gross underestimation of the
isovector charge and magnetic radii. This very fundamental constraint is very
often overlooked. In the framework of chiral pertubation theory, this 
enhancement is also present at the one--loop level as first shown in
ref.\cite{GSS}. Let me briefly explain how one can find $t_c$ by using
the so--called Landau equations (for an introduction, see e.g. the
textbook by Itzykson and Zuber~\cite{IZ}). For the pertinent one loop
graph shown in fig.~1, we have to analyze the imaginary
part. 
\begin{figure}[ht]
\label{figLan}
\hskip 2in
\epsfysize=1.8in
\epsffile{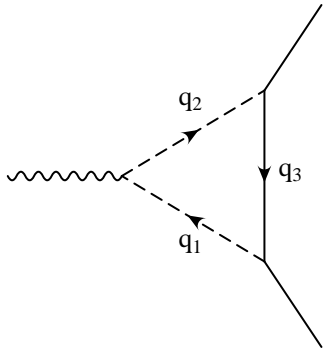}
\vspace{-0.1cm}
 \caption{The so-called triangle diagram leading to the anomalous 
          threshold in the isovector spectral functions.
      Solid, dashed and wiggly lines denote nucleons, pions and
      photons, in order.  }
\end{figure}
For that, one sets the particles in the loop on their mass
shell, $q_1^2 = q_2^2 = M_\pi^2$, $q_3^2 = m^2$. Furthermore, the loop
momenta must be linearly dependent, $\alpha_1 q_1 + \alpha_2 q_2 +
\alpha_3 q_3 = 0$ which translates into the condition that the
determinant formed from the scalar products must vanish, det($q_i
\cdot q_j$) =0 ($i,j =1,2,3$), i.e.
\beq
{m^2 \over 4} \, t\, \biggl(4 M^2_\pi - {M_\pi^4 \over m^2} - t
\biggr) = 0
\quad ,\eeq
leading to the anomalous threshold. The subdeterminant in the space
of the meson momenta ($i,j=1,2$) is given by $t(4M_\pi^2 -t)/4$ and
thus leads to the normal threshold. Furthermore, the imaginary part
of the triangle diagram shown in fig.~1, 
denoted ${\cal G}$, is readily worked out,
\beq
{\rm Im} \, {\cal G} = {1 \over 8\pi \sqrt{t(4m^2-t)}} \arctan 
{\sqrt{(t-4M_\pi^2)(4m^2-t)} \over t - 2M_\pi^2} \quad .
\eeq
At $t=t_c$, the argument of the arctan is $i$ and since $2\arctan (ix)
= i \ln[(1+x)/(1-x)]$, we recover the announced logarithmic singularity.
Recently, the question whether a similar phenomenon appears in
the isoscalar spectral function has been answered \cite{bkmff}. For that, one has
to consider two--loop graphs as shown in fig.2. 
Although the analysis of Landau equations reveals a branch point on 
the second Riemann sheet,
\beq
\sqrt{t_c} = M_\pi \, \biggl( \sqrt{4 -M_\pi^2/m^2} + \sqrt{1 -
  M_\pi^2/m^2} \biggr) \to t_c = 8.9 \,M_\pi^2 \,\,\, ,
\eeq 
close to the threshold $t_0 = 9M_\pi^2$,
the three--body  phase factors suppress its influence
in the physical region. Consequently, the spectral functions rise smoothly
up to the $\omega$ pole and the common practise of simply retaining vector
meson poles at low $t$ in the isoscalar channel is justified.
Similarly, the three--pion contribution to the nucleon isovector axial
form factor is numerically very small and can safely be neglected (as
compared e.g. to the correlated $\pi \rho$--exchange).
\begin{figure}[ht]
\label{figtwo}
\hskip 4in
\epsfysize=2.2in
%\epsffile{dis1.ps}
%\epsffile{fft1.ps}
\epsffile{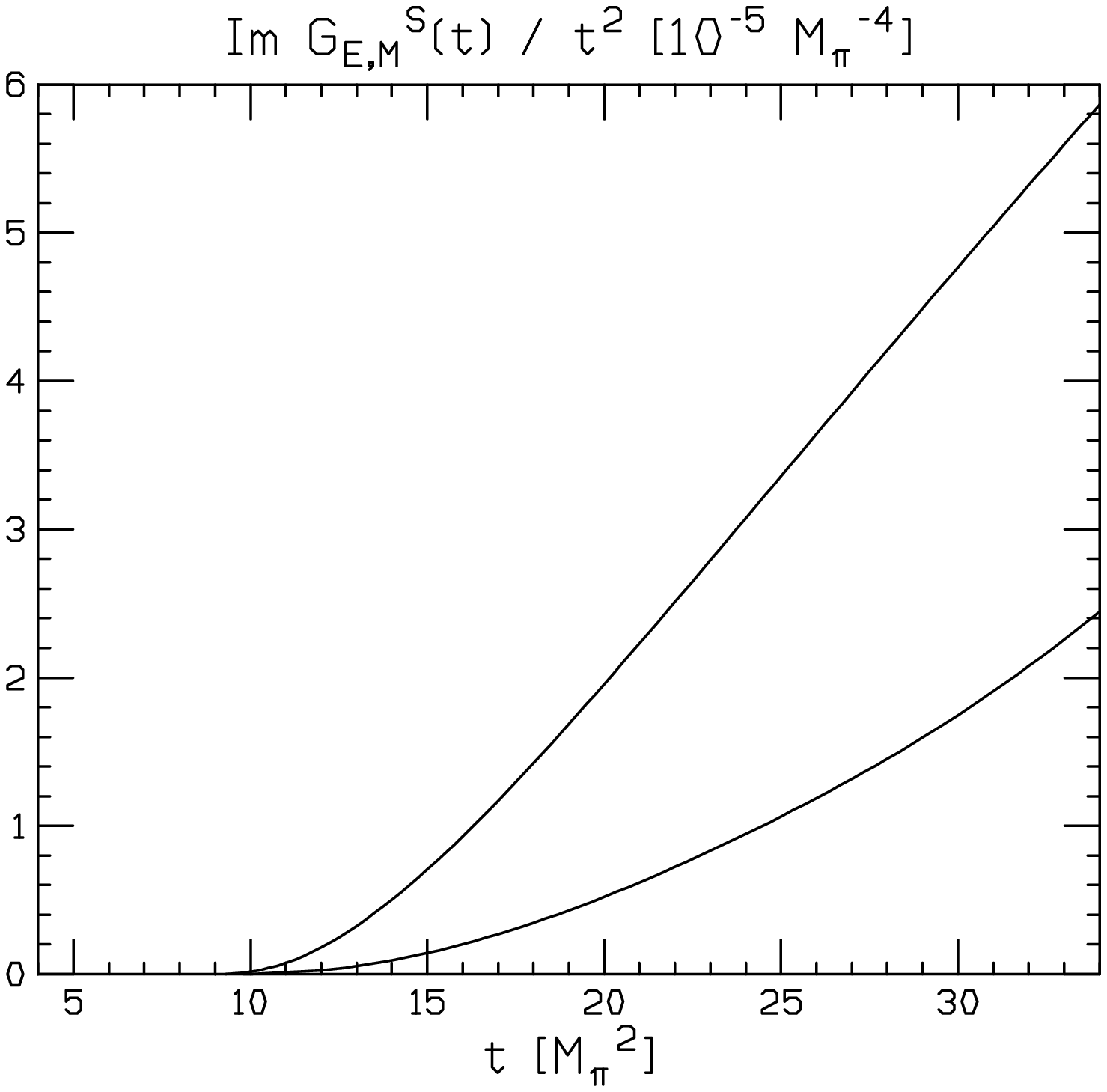}

%\protect{\hskip -1truein}

\vskip -2in
%\hskip -1truein
\epsfysize=2.in
\hskip .5truein
\epsffile{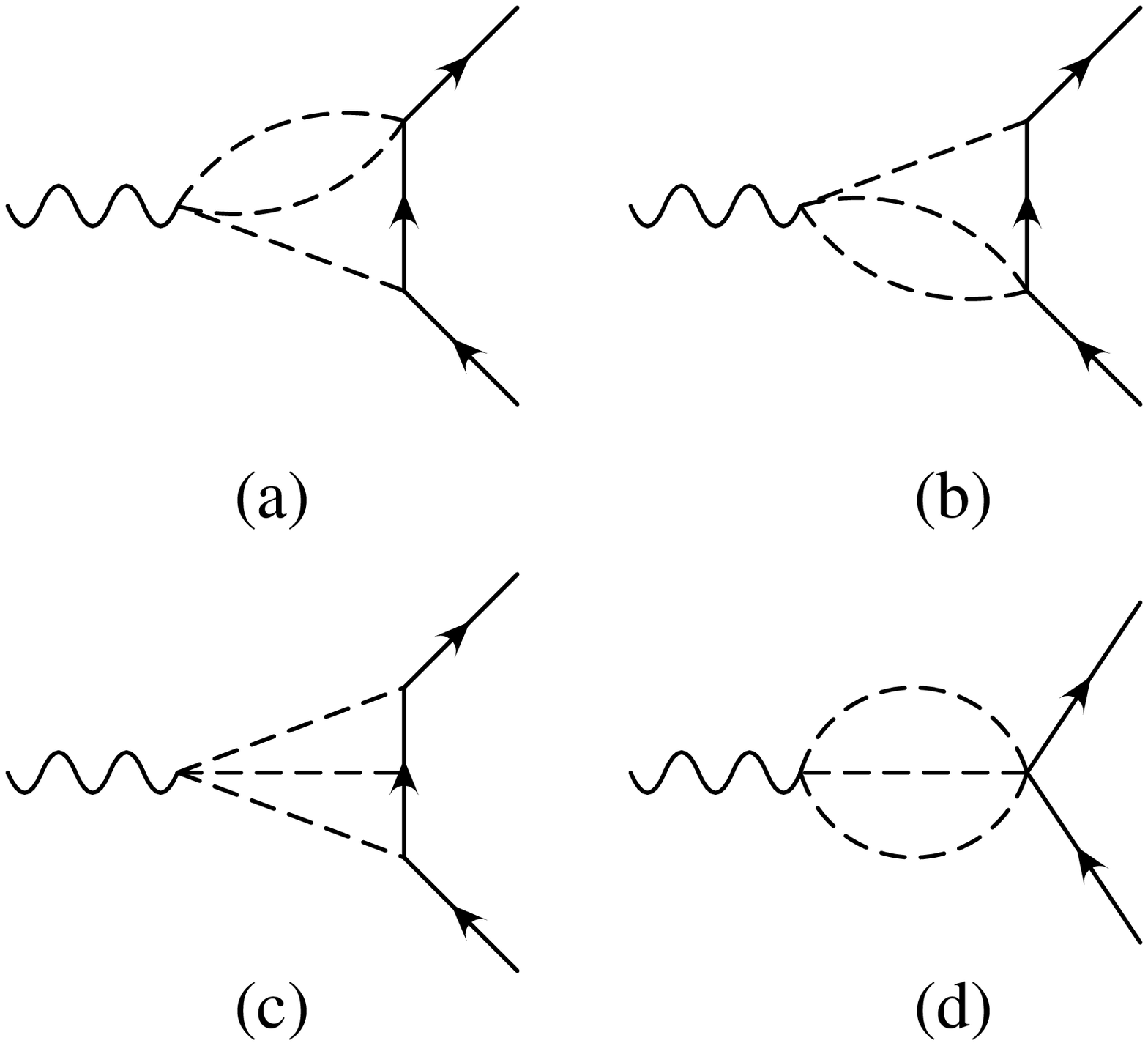}
\vspace{-0.1cm}
 \caption{
      Isoscalar spectral functions weigthed with $1/t^2$ for the electric
      (lower) and magnetic (upper line) Sachs ff (right panel). 
      In the left panel the underlying two--loop graphs are shown (solid,
      dashed, wiggly lines: Nucleons, pions and photons).
    \label{exfig} }
\end{figure}

%%%%%%%%%%%%%%%%%%%%%%%%%%%%%%%%%%%%%%%%%%%%%%%%%%%%%%%%%%%%%%%%%%%%%%%%%%
\section{Confronting the data}
\label{sec:data}
\setcounter{equation}{0}

As already stated in the introduction, most of the precision data on
the nucleon at low energies come from processes involving real or
virtual photons such as pion photo-- and electroproduction as well as
Compton scattering. This is mostly due to the advent of the CW
accelerators such as MAMI at Mainz and improved detector
technology. It is worth to stress that there are on--going activities
in these fields also at LEGS (Brookhaven), SAL (Saskatoon), BATES
(MIT), ELSA (Bonn) and other places. Clearly, CEBAF at Jefferson Lab
will significantly contribute accurate data to a large variety of processes.
Other precise data come from
atomic energy shifts at PSI (Villigen)   and there is also a
tremendous amount of threshold data for $\pi N \to \pi \pi N$ from such
places like TRIUMF (Vancouver). Just from the beginning I would like to
stress that one should not see these different experiments and data in
isolation but that they are rather intimately connected. For example,
the imaginary parts of the various multipoles in pion photo-- and
electroproduction are proportional to the respective pion--nucleon
scattering phase shifts via the Fermi--Watson final state theorem. 
%As
%an example, let me give the imaginary part of the electric dipole
%amplitude in neutral pion photoproduction off protons,
%\beq {\rm Im}~E_{0+}^{\pi^0 p} = {\rm Re}~E_{0+}^{(0)} \tan(\delta_{1/2}) +
%\frac{1}{3}{\rm Re}~E_{0+}^{(1/2)} \tan(\delta_{1/2}) +  
%\frac{2}{3}{\rm Re}~E_{0+}^{(3/2)} \tan(\delta_{3/2})  \, \, , \eeq
%with $\delta_{1/2,3/2}$ the pion--nucleon S--wave phase shifts for
%total isospin 1/2 and 3/2, respectively. 
Furthermore, various contact terms
show up in different reactions. For example, the LECs $c_{1,2,3}$
which can be determined in $\pi N \to \pi N$ \cite{bkmpin} show up in
the order $p^4$ calculation of the nucleons' electromagnetic
polarizabilities \cite{bkms} from non--vanishing insertions of ${\cal
  L}_{\pi N}^{(2)}$. This should always be kept in mind. For a nice
and instructive flow--chart giving the links between low--energy pion 
experiments I refer the reader to Fig.7.33--1 in the monograph of 
de~Benedetti \cite{debe}. A remark on the pion--nucleon coupling 
constant is in order. In most of the calculations shown, the 
Karlsruhe--Helsinki value of $g_{\pi N}= 13.4$ was used. There seem,
however, to be indications from various sources that the value is
smaller,  $g_{\pi N}= 13.05$. This can be accounted for by simply
plugging in this value into the formulae given in the respective
papers. I will come back to this point when discussing the so--called
Goldberger--Treiman discrepancy. For the illustrative purpose of these
lectures, however, the results based on the larger value of $g_{\pi N}$
are perfectly suitable.

\subsection{Charged and neutral pion photoproduction}

\noindent Charged pion photoproduction at threshold is well described
in terms of the Kroll--Ruderman contact term, which is non--vanishing
in the chiral limit. All chiral corrections including the
third order in the pion mass have been calculated \cite{bkmcp}. The
chiral series is quickly converging and the theoretical error on the
CHPT predictions is rather small, see table~\ref{tab:char}. The LECs have been
determined from resonance exchange. 
Notice that these uncertainties do not account for the variations in 
pion--nucleon coupling constant.
The available threshold data are quite  old, with
the exception of the recent TRIUMF experiment on the inverse reaction
$\pi^- p \to \gamma n$ and the SAL measurement for $\gamma p \to \pi^+ n$.
While the overall agreement is quite good for 
the $\pi^+ n$ channel, in the $\pi^- p$ channel the CHPT prediction
is on the large side of the data. Clearly, we need more precise data
to draw a final conclusion. It is, however, remarkable to have
predictions with an error of only 2\%$\,$.
\renewcommand{\arraystretch}{1.4}
\begin{table}[hbt]
\begin{center}
\begin{tabular}{|l|c|c|c|}
\hline
                                & CHPT\protect{\cite{bkmpi0}}
                                & Order  & Experiment  \\
\hline
$E_{0+}^{\rm thr} (\pi^+ n)$    & $28.2 \pm 0.6$ & $p^4$  
                                & $27.9\pm 0.5$\protect{\cite{burg}},
                                $28.8 \pm 0.7$\protect{\cite{adam}},
                                $27.6 \pm 0.6$\protect{\cite{salp}}   \\
$E_{0+}^{\rm thr} (\pi^- p)$    & $-32.7 \pm 0.6$  & $p^4$ & $-31.4
                                \pm 1.3$\protect{\cite{burg}}, 
                               $-32.2 \pm 1.2$\protect{\cite{gold}}, 
                               $-31.5\pm 0.8$\protect{\cite{triumf}}  \\
\hline
\end{tabular}
\caption{Predictions and data for the charged pion 
electric dipole amplitudes in $10^{-3}/M_{\pi^+}$.\label{tab:char}}
\end{center}
\end{table}
\noindent The threshold production of neutral pions is much more
subtle since the corresponding electric dipole amplitudes vanish in
the chiral limit. Space does not allow to tell the tale of the
experimental and theoretical developments concerning the electric
dipole amplitude for neutral pion production off protons, for details
see \cite{ulf95}. Even so the convergence for this particular
observable is slow, a CHPT calculation to order $p^4$ does allow to
understand the energy dependence of $E_{0+}$ in the threshold region
once three LECs are fitted to the total and differential cross section
data~\cite{bkme0p}. The threshold
value agrees with the data, see table~\ref{tab:e0plus0}.
More interesting is the
case of the neutron. Here, CHPT predicts a sizeably larger $E_{0+}$
than for the proton (in magnitude).
The CHPT predictions for $E_{0+} (\pi^0 p,n)$ in the threshold
region  clearly
exhibit the unitary cusp due to the opening of the secondary
threshold, $\gamma p \to \pi^+ n \to \pi^0 p$ and $\gamma n \to \pi^-
p \to \pi^0 n$, respectively. Note, however, that while $E_{0+} (\pi^0
p)$ is almost vanishing after the secondary threshold, the neutron
electric dipole amplitude is sizeable ($-0.4$ compared to $2.8$ in
units of $10^{-3}/M_{\pi^+}$).
The question arises how to measure the neutron amplitude?
The natural neutron target is the deuteron. The corresponding electric
dipole $E_d$ amplitude has been calculated to   order $p^4$ in
ref.\cite{bblmvk}. It was shown that the next--to--leading order
three--body corrections and the possible four--fermion contact terms
do not induce any new unknown LEC and one therefore can calculate
$E_d$ in parameter--free manner. Furthermore, the leading order
three--body terms are dominant, but one finds a good convergence for
these corrections and also a sizeable sensitivity to the elementary
neutron amplitude. Note also that neutral pion production off
deuterium has recently been measured at SAL. Still, there is more
interesting physics in these channels.
Quite in contrast to what was believed for a long
time, there exist a set of LETs for the slopes of the P--waves
$P_{1,2} = 3E_{1+} \pm M_{1+} \mp M_{1-}$ at threshold, e.g.
\beq \frac{1}{|\vec q \,|} P_{1, {\rm thr}}^{\pi^0 p} = \frac{e g_{\pi
  N}}{8 \pi m^2} \left\lbrace 1 + \kappa_p + \mu \left[ -1 - 
\frac{\kappa_p}{2} + \frac{g_{\pi N}^2}{48 \pi}(10 -3\pi) \right]
+ {\cal O}(\mu^2) \right\rbrace \, \, . \eeq
Numerically, this translates into 
\beq \frac{1}{|\vec q \,|} P_{1, {\rm thr}}^{\pi^0 p} = 0.512 \, ( 1
-0.062) \, {\rm GeV}^{-2} = 0.480 \, {\rm GeV}^{-2} \, \, , \eeq
which is given in table~\ref{tab:e0plus0} 
in units which are more used in the literature. The agreement with
the data is stunning. A theoretical uncertainty on this order $p^3$ 
calculation can only be given when the next order has been calculated.
That calculation is underway. Soon, there will also be an experimental value
for $P_2$ at threshold once the MAMI data on $\vec{\gamma} p \to \pi^0
p$ have been analyzed.
\renewcommand{\arraystretch}{1.4}
\begin{table}[hbt]
\begin{center}
\begin{tabular}{|l|c|c|c|c|}
\hline
                        & CHPT & Order & Experiment & Units  \\
\hline
$E_{0+}^{\rm thr} (\pi^0 p)$    & $-1.16$\protect{\cite{bkme0p}} & $p^4$ 
                                & $-1.31\pm 0.08$\protect{\cite{fuchs}},
                                $-1.32\pm 0.05$\protect{\cite{berg}}
                                & $10^{-3}/M_{\pi^+}$ \\
$P_{1}^{\rm thr} (\pi^0 p)$     & $10.3$\protect{\cite{bkmcp}}  & $p^3$ 
                                & $10.02\pm 0.15$\protect{\cite{fuchs}},
                                $10.26\pm 0.1$\protect{\cite{berg}}
                                & $|q||k|10^{-3}/M_{\pi^+}^3$  \\
$E_{0+}^{\rm thr} (\pi^0 d)$    & $-1.8\pm0.2$\protect{\cite{bblmvk}}  
                                & $p^4$ & 
                                $-1.7 \pm 0.2$\protect{\cite{argan}}, 
                                $-1.5 \pm 0.1$\protect{\cite{sald}} 
                                & $10^{-3}/M_{\pi^+}$  \\
 \hline
\end{tabular}
\caption{Predictions and data for  neutral pion S-- and P--wave
         multipoles.\label{tab:e0plus0}}
\end{center}
\end{table}

\subsection{Pion electroproduction}
Producing the pion with virtual photons offers further insight
since one can extract the longitudinal S--wave multipole $L_{0+}$ and
also novel P--wave multipoles. Data have been taken at
NIKHEF~\cite{welch}\cite{benno} and MAMI~\cite{distler} for
photon virtuality of $k^2 = -0.1$~GeV$^2$. In fact, it has been argued
previously that such photon four--momenta are already too large for
CHPT tests since the loop corrections are large~\cite{bklm}. However,
these calculations were performed in relativistic baryon CHPT and thus
it was necessary to redo them in the heavy fermion formalism. This was
done in~\cite{bkmel}. The abovementioned data for differential cross
sections were used to determine the three novel S--wave LECs. I should
mention that one of the operators used is of dimension five, i.e. one
order higher than the calculation was done. This can not be
circumvented since it was shown that the two S--waves are
overconstrained by a LET valid up to order $p^4$. The resulting
S--wave cross section $a_0 = |E_{0+}|^2 + \ve_L \,|L_{0+}|^2$ shown in  
fig.~3 is in fair agreement with the data. Note also that it is
dominated completely by the $L_{0+}$ multipole (upper dot-dashed line)
since $E_{0+}$ passes through zero at $k^2 \simeq
-0.04$~GeV$^2$. However, in agreement with the older (and less precise)
calculations, the one loop corrections are large so one should compare at
lower photon virtualities. In ref.\cite{bkmel}, many predictions for
$k^2 \simeq -0.05$~GeV$^2$ are given. At MAMI, data have been taken in
this range of $k^2$ and we are looking forward to their analysis, in
particular it will be interesting to nail down the zero--crossing of
the electric dipole amplitude and to test the novel P--wave 
LETs~\cite{bkmprl}.
\begin{figure}[hbt]
\hskip 1.3in
\epsfysize=2.5in
\epsffile{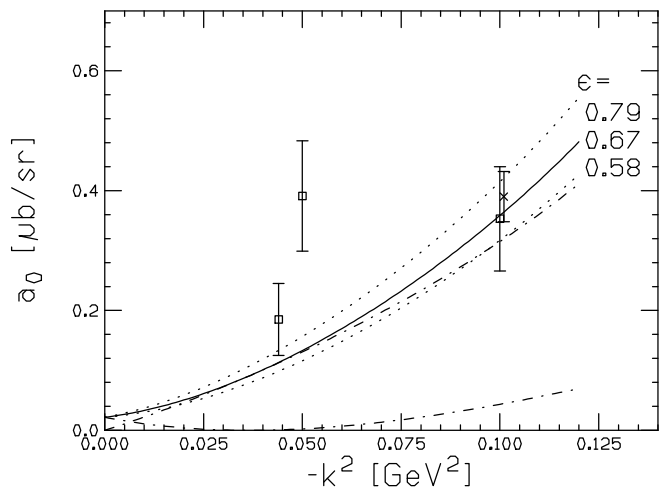}
\vspace{-0.1cm}
 \caption{The S--wave cross section $a_0$ compared to the data for
          various photon polrizations. The upper (lower) dash--dotted
          line gives the contribution of the longitudinal (electric)
          dipole amplitude to $a_0$ for $\epsilon = 0.79$ (solid
          line).  }
\end{figure}%%\label{figao}

\subsection{Compton scattering}

As discussed before, the low--energy Compton scattering amplitude
can be decomposed in a spin--independent and a spin--dependent part.
In the former case, the nucleon structure is encoded in the so--called
electromagnetic polarizabilities.  These have been measured over the years at
Illinois, Mainz, Moscow, Oak Ridge and Saskatoon. 
Calculations have been performed to fourth order with some LECs
determined from data and others from resonance saturation. A summary
is given in table~\ref{tab:co}. While in case of the
proton a consistent picture is emerging~\cite{bkms}, the only published empirical
values deduced from scattering slow neutrons on heavy atoms have
recently been put into question~\cite{koes}. This experimental problem remains to
be sorted out. The most promising result is the almost complete 
cancellation of a negative  non--analytic pion loop contribution with the large
positive contribution of tree level $\Delta$--exchange in case of the
protons magnetic polarizability~\cite{bkms}. This is the first time that a
consistent picture of the para-- and diamagnetic contributions to
$\bar{\beta}_p$ has been found and it underlies the importance of chiral,
i.e. pion loop, physics in understanding the nucleon structure as 
revealed in Compton scattering. Furthermore, there exist chiral 
predictions for the
spin--dependent polarizabilities. As of today, no direct measurements
exist but some indirect information based on multipole analysis points
towards the important role of the $\Delta (1232)$ as an active dof to
understand these quantities. I refer to the Mainz chiral dynamics proccedings 
\cite{mzproc} for a more detailed discussion. 
 
\renewcommand{\arraystretch}{1.3}
\begin{table}[hbt]
\begin{center}  
\begin{tabular}{|c|c|c|c|c|c|c|} \hline
  & Prediction &  Order  & Ref. & Data & Ref. & Units  \\ 
\hline
$\bar{\alpha}_p$ & $10.5\pm 2.0$  & $p^4$& \cite{bkms} & $10.4\pm0.6$ 
& \cite{feder} \cite{sal}\cite{macg} & 10$^{-4}$ fm$^{3}$  \\ 
$\bar{\beta}_p$ & $3.5\pm 3.6$  & $p^4$& \cite{bkms} & $3.8\mp0.6$ 
& \cite{feder} \cite{sal}\cite{macg} & 10$^{-4}$ fm$^{3}$  \\
$\bar{\alpha}_n$ & $13.4\pm 1.5$  & $p^4$& \cite{bkms} & $12.3\pm1.3$ 
& \cite{schmi} & 10$^{-4}$ fm$^{3}$  \\
$\bar{\beta}_n$ & $7.8\pm 3.6$  & $p^4$& \cite{bkms} & $3.5\mp 1.3$ 
& \cite{schmi} & 10$^{-4}$ fm$^{3}$  \\ \hline
%$ g_P $ & $ 8.44 \pm 0.23$  & $q^3$& \cite{bkmgp} &
%$8.7\pm 1.6$  & \cite{richter} & --  \\ & & & & & & 
% \\ \hline 
\end{tabular} 
\end{center}
\caption{Chiral predictions for 
the electric ($\bar \alpha$) and
magnetic ($\bar \beta$) polarizabilities of the proton and the
neutron. Note that a recent reanalysis of the Oak Ridge experiment
does not lead to the same values as the published ones[100].\label{tab:co}}
\end{table}

\subsection{Pion--nucleon scattering}

There is, of course, ample of data on elastic pion--nucleon scattering
in the threshold region. The one-loop contribution to the 
$\pi N$-scattering amplitude 
to order $p^3$ in HBCHPT has first been worked out by Moj\v zi\v
s~\cite{moj}. Here, I 
follow ref.\cite{bkmlec} in which certain aspects of 
pion--nucleon scattering have 
also been addressed. In the center-of-mass frame the $\pi
N$-scattering amplitude $\pi^a(q) + N(p) \to \pi^b(q') + N(p')$ takes the
following form: 
\begin{equation} 
T^{ba}_{\pi N} = \delta^{ba} \Big[ g^+(\omega,t)+ i \vec
\sigma \cdot(\vec q\,'\times \vec q\,) \, h^+(\omega,t) \Big] +i \epsilon^{bac}
\tau^c \Big[ g^-(\omega,t)+ i \vec \sigma \cdot(\vec q\,'\times \vec q\,) \,
h^-(\omega,t) \Big] 
\end{equation}
with $\omega = v\cdot q = v\cdot q\,'$ the pion cms energy and $t=(q-q\,')^2$ 
the invariant momentum transfer squared. $g^\pm(\omega,t)$ refers to the
isoscalar/isovector non-spin-flip amplitude and $h^\pm(\omega,t)$ to the
isoscalar/isovector spin-flip amplitude. After renormalization of the pion
decay constant $F_\pi$ and the pion-nucleon coupling constant $g_{\pi N}$, one
can give the one-loop contributions to the cms amplitudes
$g^\pm(\omega,t)$ and $h^\pm(\omega,t)$ at order $p^3$ in closed form, 
see ref.\cite{bkmlec}.
In table~\ref{tab:piN}, I show the predictions for the remaining S, P, D and
F-wave threshold parameters which were not used in the fit to determine the
LECs. In some cases, contributions from the dimension three Lagrangian appear.
The corresponding LECs have been estimated using resonance exchange. In 
particular, the 10\% difference in the P--wave scattering volumina $P_1^-$
and $P_2^+$ is a clear indication of chiral loops, because
nucleon and $\Delta$ Born
terms give the same contribution to these two observables. Note also that
the eight D-- and F--wave threshold parameters to this order are free of
contributions from dimension three and thus uniquely predicted. The overall
agreement of the predictions with the existing experimental values is rather
satisfactory.  Still, a complete calculation to next order is called for.
\renewcommand{\arraystretch}{1.3}
\begin{table}[t]
\begin{center}
\begin{tabular}{|c|c|c|c|c|c|c|}
    \hline  \hline
    Obs. & CHPT &  Order & Ref. & Exp. value & Ref. & Units \\
    \hline
$a^-$ &  $9.2\pm 0.4$ & $p^4$ & \protect{\cite{bkmprc}}
& $ 8.4 \ldots 10.4$ &  \protect{\cite{sigg}}
& $10^{-2}\, M_\pi^{-1} $ \\
$b^-$ &  $2.01$ & $p^3$ & \protect{\cite{bkmlec}} & $ 1.32 \pm 0.62$ &
\protect{\cite{hoeh}} & $10^{-2}\, M_\pi^{-1} $ \\
   \hline
$P_1^-$ &  $-2.44\pm 0.13$ & $p^3$ & \protect{\cite{bkmlec}}
& $ -2.52 \pm 0.03$ &
\protect{\cite{hoeh}} & $ M_\pi^{-3} $ \\
$P_2^+$ &  $-2.70\pm 0.12$ & $p^3$ & \protect{\cite{bkmlec}} & $ -2.74 \pm 0.03$ &
\protect{\cite{hoeh}} & $ M_\pi^{-3} $ \\
   \hline
$a^+_{2+}$ &  $-1.83$ & $p^3$ & \protect{\cite{bkmlec}} & $ -1.8 \pm 0.3$ &
\protect{\cite{hoeh}} & $10^{-3}\, M_\pi^{-5} $ \\
$a^+_{2-}$ &  $2.38$ & $p^3$ & \protect{\cite{bkmlec}} & $ 2.20 \pm 0.33$ &
\protect{\cite{hoeh}} & $10^{-3}\, M_\pi^{-5} $ \\
$a^-_{2+}$ &  $3.21$ & $p^3$ & \protect{\cite{bkmlec}} & $ 3.20 \pm 0.13$ &
\protect{\cite{hoeh}} & $10^{-3}\, M_\pi^{-5} $ \\
$a^-_{2-}$ &  $-0.21$ & $p^3$ & \protect{\cite{bkmlec}} & $ 0.10 \pm 0.15$ &
\protect{\cite{hoeh}} & $10^{-3}\, M_\pi^{-5} $ \\
   \hline
$a^+_{3+}$ &  $0.29$ & $p^3$ & \protect{\cite{bkmlec}} & $ 0.43 $ &
\protect{\cite{hoeh}} & $10^{-3}\, M_\pi^{-7} $ \\
$a^+_{3-}$ &  $0.06$ & $p^3$ & \protect{\cite{bkmlec}} & $ 0.15 \pm 0.12$ &
\protect{\cite{hoeh}} & $10^{-3}\, M_\pi^{-7} $ \\
$a^-_{3+}$ &  $-0.20$ & $p^3$ & \protect{\cite{bkmlec}} & $ -0.25 \pm 0.02$ &
\protect{\cite{hoeh}} & $10^{-3}\, M_\pi^{-7} $ \\
$a^-_{3-}$ &  $0.06$ & $p^3$ & \protect{\cite{bkmlec}} & $ 0.10 \pm 0.02$ &
\protect{\cite{hoeh}} & $10^{-3}\, M_\pi^{-7} $ \\
   \hline   \hline
  \end{tabular} \end{center}
\caption{Threshold parameters predicted by CHPT. The order
of the prediction is also given together with the "experimental
values". Only $a^\pm$ can be measured from pionic atoms. The other
values come from the Karlsruhe--Helsinki dispersive analysis.
\label{tab:piN}}
\end{table}
In ref.\cite{bebeleme}, it was shown that  pion scattering off
deuterium can give some bounds on the isoscalar $\pi N$ scattering
length and therefore a particular combination of the LECs $c_1,c_2$
and $c_3$. The resulting value is consistent with the one found before.
The formalism spelled out in section~\ref{sec:vir} has been applied
in ref.\cite{suiso} to calculate the isospin violating corrections
to neutral pion scattering off protons and neutrons to  ${\cal
  O}(p^3)$. It substantiates Weinberg's claim made in
1977~\cite{weinmass} that there are sizeable effects of isospin
violation, stemming in part from the light quark mass difference and
also from electromagnetism.

\subsection{The reaction $\pi N \to \pi\pi N$}

Single pion production off nucleons has been at the center of numerous
experimental and theoretical investigations since many years. One of
the original motivations of these works was the observation that
the elusive pion--pion threshold S--wave interaction could be deduced
from the pion--pole graph contribution. A whole series of precision
experiments at PSI, TRIUMF and CERN has been performed over the last
decade and there is still on--going activity. On the theoretical side,
chiral perturbation theory has been used to consider these processes.
Beringer considered the reaction $\pi N \to \pi
\pi N$ to lowest order in chiral perturbation theory
\cite{bering}. Low--energy theorems for the threshold amplitudes $D_1$
and $D_2$ were derived in \cite{bkmplb},
\beq D_1 = \frac{g_A}{8 F_\pi^3} \left( 1 +\frac{7 M_\pi}{2m} \right)
+ {\cal O}(M_\pi^2)  = 2.4 \, \, {\rm fm}^3 \,\, , \label{lppn1}\eeq 
\beq D_2 = -\frac{g_A}{8 F_\pi^3} \left( 3 +\frac{17 M_\pi}{2m} \right)
+ {\cal O}(M_\pi^2)  = -6.8 \, \, {\rm fm}^3 \,\,  . \label{lppn2} \eeq
These are free of unknown parameters and not sensitive to the $\pi
\pi$--interaction beyond tree level. 
A direct comparison with the threshold data for
the channel $\pi^+ p \to \pi^+ \pi^+ n$, which is only sensitive to $D_1$,
leads to a very satisfactory description whereas in case of the process
$\pi^- p \to \pi^0 \pi^0 n$, which is only sensitive to $D_2$, sizeable 
deviations are found for the total cross sections near threshold. These were 
originally attributed to the strong pionic final--state interactions in the
$I_{\pi\pi}=0$ channel. However, this conjecture turned out to be incorrect 
when a complete higher order calculation of the threshold amplitudes $D_{1,2}$
was performed \cite{bkmppn}. In that paper, the relation
between the threshold amplitudes $D_1$ and $D_2$ for the reaction $\pi N \to
\pi \pi N$ and the $\pi\pi$ S--wave scattering lengths $a_0^0$ and $a_0^2$
in the framework HBCHPT to second order in the pion mass was worked
out (for details, I refer to that paper). 
Notice that the pion loop and pionic counterterm 
corrections  only start  contributing to the $\pi\pi N$ threshold amplitudes at
second order. One of these counterterms, proportional to the low--energy
constant $\ell_3$, eventually allows to
measure the scalar quark condensate, i.e. the strength of the spontaneous 
chiral symmetry breaking in QCD. However, at that order, the largest 
contributions to $D_{1,2}$ stem indeed from insertions of the dimension two
chiral pion--nucleon Lagrangian, which is characterized by the LECs
constants called $c_i$.  In particular this is the case for the amplitude
$D_2$. To be  specific, consider the threshold  amplitudes $D_{1,2}$ calculated
from the relativistic Born graphs (with lowest order vertices) and the
relativistic $c_i$--terms expanded  to second order in the pion mass. This 
gives 
\beqa \label{D12}
D_1^{{\rm Born}} + D_1^{c_i} &=& (2.33 + 0.24 \pm 0.10) \, {\rm
    fm}^3 = (2.57 \pm 0.10) \, {\rm  fm}^3 \,\,, \\
D_2^{{\rm Born}} + D_2^{c_i} &=& (-6.61 - 2.85 \pm 0.06) \, {\rm
    fm}^3 = (-9.46 \pm 0.06) \, {\rm  fm}^3 \,\, ,
\eeqa
which are within 14\% and 5\% off the empirical values, 
\beq D_1^{\exp} =(2.26 \pm
0.10)\,{\rm fm}^3 \,\,\, , \quad D_2^{\exp} = (-9.05 \pm 0.36)\, {\rm
  fm}^3 \,\,\, , \eeq
respectively.  It appears therefore natural to extend the same
calculation above threshold and to compare to the large body of data
for the various reaction channels that exist~\cite{bkmppn2}. 
It was already shown by
Beringer \cite{bering} that taking simply the relativistic Born terms
does indeed not suffice to describe the total cross section data for
incoming pion energies up to 400~MeV in most channels. Such a failure 
can be expected from the threshold expansion of $D_2$, where the Born terms 
only amount to 73\% of the empirical value. We therefore expect that the
inclusion of the dimension two operators, which clearly improves the
prediction for $D_2$ at threshold, will lead to a better description of
the above threshold data. In particular, it will tell to which extent
loop effects are necessary (and thus testing the sensitivity to the
pion--pion interaction beyond tree level) and to which extent one has to
incorporate explicit resonance degrees of freedom like the Roper and the
$\Delta$--isobar as well as heavier mesons ($\sigma, \rho,\omega $) 
as dynamical degrees of freedom (as it is done in
many models, see e.g. \cite{oset} \cite{jaeck}). 
Since the LECs $c_i$ have previously been determined, 
all our results to this order  are based on a 
truly parameter--free calculation.
One finds that (a) for pion energies up to $T_\pi = 250\,$MeV, in all but
one case the inclusion of the contribution $\sim c_i$ clearly improves
the description of the total cross sections (solid versus dashed line), 
most notably in the
threshold region for $\pi^- p \to \pi^0 \pi^0 n$. Up to $T_\pi =
400\,$MeV, the trend of the data can be described although some
discrepancies particularly towards the higher energies persist,
and  (b) double differential cross sections for 
$\pi^- p \to \pi^+ \pi^- n$ at incident pion energies
below $T_\pi = 250\,$MeV are well described. In fig.~4, the total
cross section for $\pi^- p \to \pi^+ \pi^- n$ is shown in comparison
to the
data~\cite{data}.
%\cite{bj80}\cite{jo74}\cite{bl70}\cite{sa70}\cite{ba65}
%\cite{bl63}\cite{de61}\cite{pe60}.
In particular, the novel TRIUMF threshold data~\cite{martin}
between threshold and $T_\pi$ show an excellent agreement with the
previously published prediction.
\vspace{-0.6in}  %%% may be -0.7 sven - grrrrrr

\begin{figure}[ht]
\hskip 3.2in
\epsfysize=4.0in
\epsffile{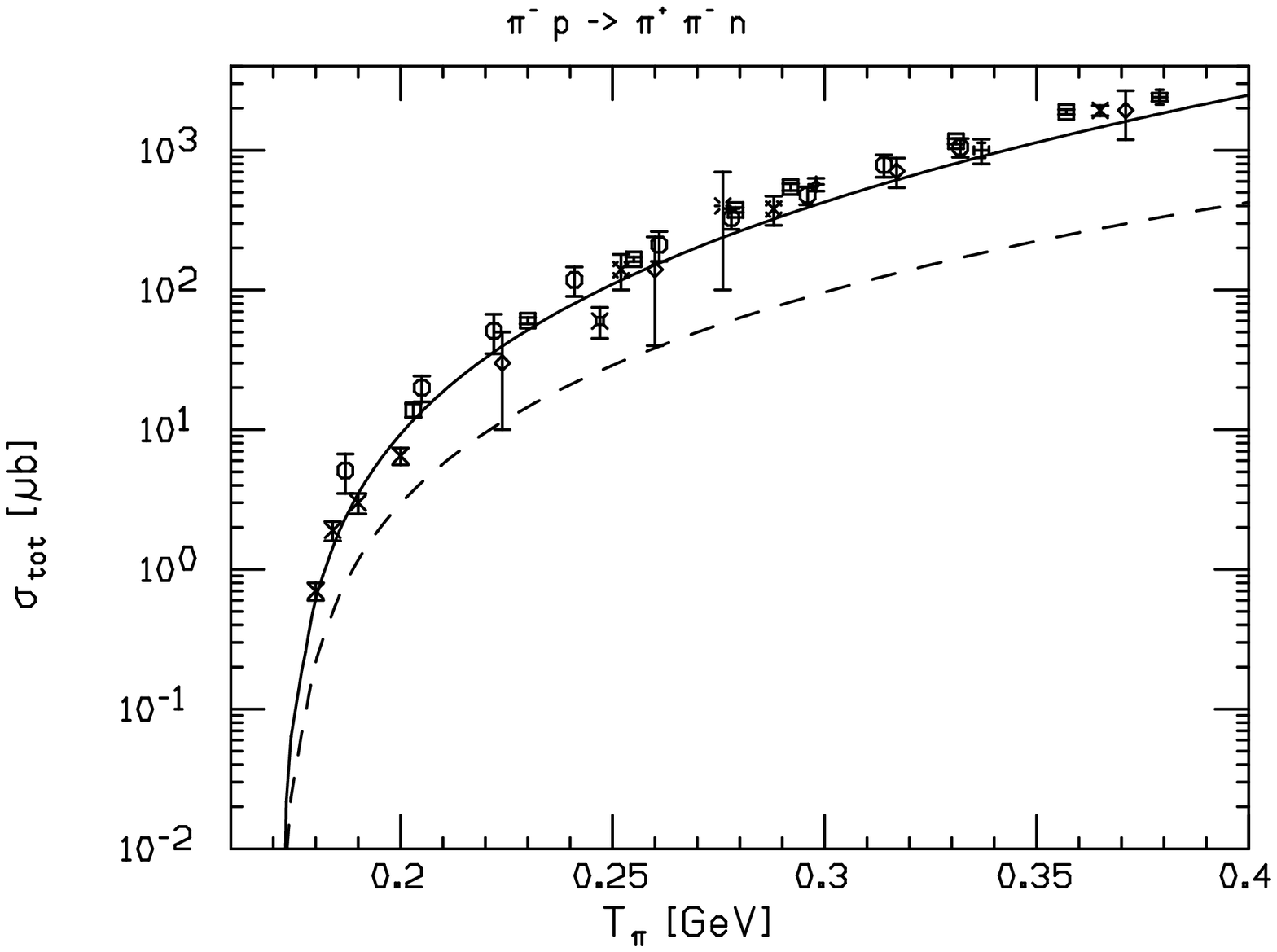}

%\protect{\hskip -1truein}

\vskip -3.2in
%\hskip -1truein
\epsfysize=2.4in
\hskip .2truein
\epsffile{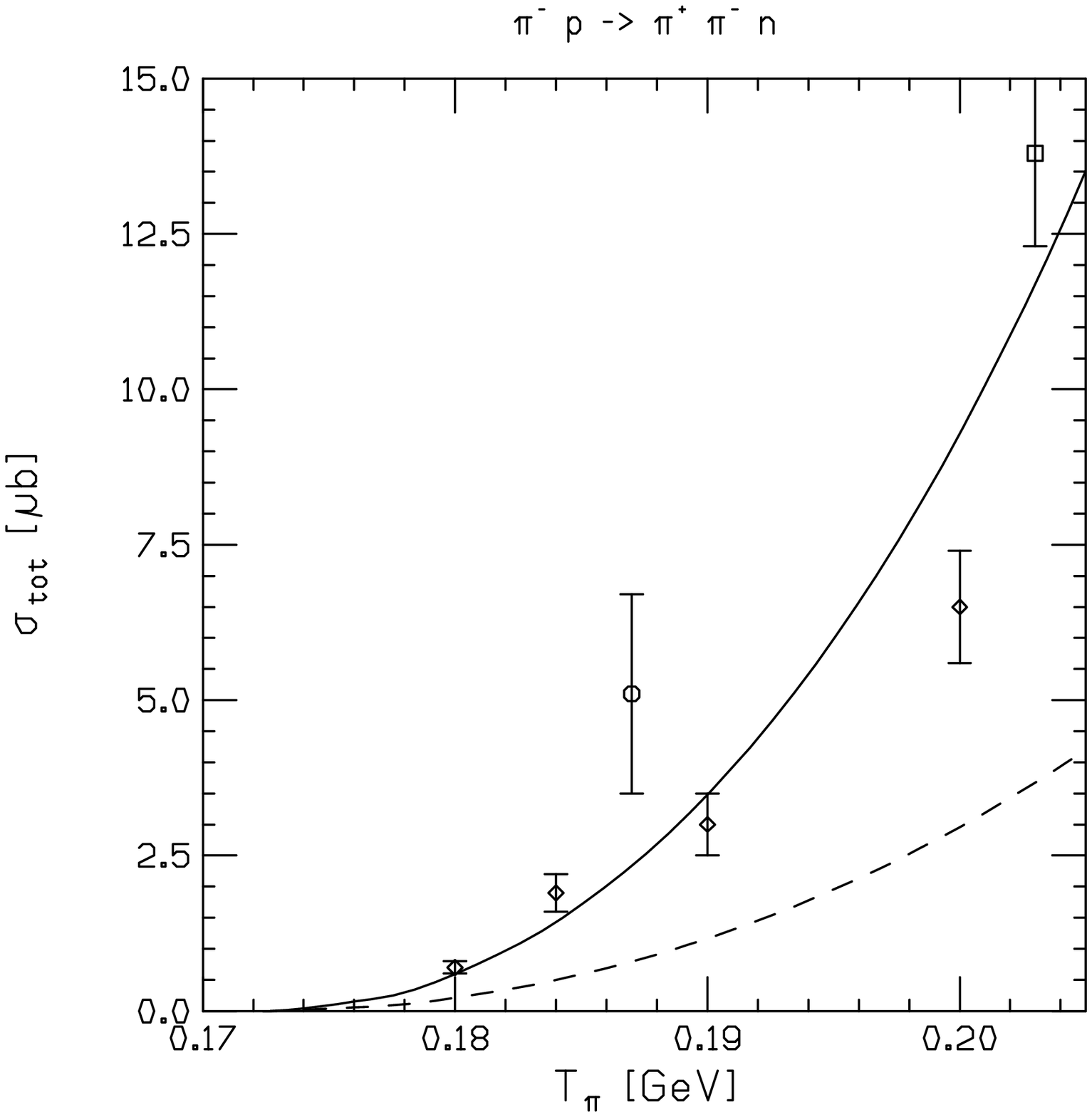}
\vspace{-0.1cm}
 \caption{Total cross section for $\pi^- p \to \pi^+ \pi^- n$. Left
   panel: The threshold region. The new TRIUMF data are depicted by
   the diamonds and have significantly smaller error bars than the
   older data. Right panel: Comparison to the data 
   up to pion kinetic energy of 400~MeV.}
\end{figure}

%\vfill

\subsection{The Goldberger--Treiman discrepancy revisited}

I briefly want to return to the Goldberger--Treiman relation,
Eq.(\ref{gtr}). In nature, it is not exactly fulfilled and one
considers the so--called Goldberger--Treiman discrepancy (GTD),
\beq
\Delta_{\pi N} \equiv 1 - {m \, g_A \over F_\pi \, g_{\pi N}} \quad.
\eeq
In fact, at this point the precise values of the axial--vector
coupling, the pion decay and, in particular, the pion--nucleon 
coupling constant come into play. To third order, the GTD is given
in terms of the LEC $b_{23}$ (see \cite{bkmrev} for a precise definition),
\beq  
\Delta_{\pi N} = - {M_\pi^2 \over 8 \pi^2 F_\pi^2} \, b_{23} \,\,\, .
\eeq
Furthermore, if one describes the whole GTD by a form factor effect
and chooses one particular form for the pion interpolating field,
one can translate the value of $b_{23}$ into the cut--off of a
monopole pion--nucleon form factor via (for details, see \cite{bkmrev})
\beq
\Lambda = {4 \pi F_\pi \over \sqrt{-2\,b_{23}}} \quad.
\eeq 
In table~8, I have summarized the status for various input values. The
first two values for $g_A$ refer to the PDG 1996 value~\cite{PDG96} 
and the next two to the most recent Grenoble measurement~\cite{gren}.
I use throughout the PDG value for $F_\pi$ and give results for the
two values of $g_{\pi N}$ discussed in the introduction of this
section. It should be noted that for the smaller value of $g_{\pi N}$,
which seems to be most widely accepted, the GTD is less than 2\% and
the resulting form factor is no longer as soft as claimed previously.

\renewcommand{\arraystretch}{1.3}
\begin{table}[hbt]
\begin{center}
\begin{tabular}{|c|c|c|c|c|c|}
    \hline  \hline
    $g_A$&$F_\pi$ [MeV] &$g_{\pi N}$ & $\Delta_{\pi N}$ [\%] & $b_{23}$ 
    & $\Lambda$ [MeV] \\
    \hline
%    1.260 & 93.00 & 13.40 & 5.1 & -1.80 & 616 \\
%    1.260 & 93.00 & 13.05 & 2.6 & -0.91 & 869 \\
    1.260 & 92.42 & 13.40 & 4.5 & -1.57 & 656 \\
    1.260 & 92.42 & 13.05 & 2.0 & -0.68 & 994 \\
%    1.266 & 93.00 & 13.40 & 4.7 & -1.64 & 645 \\
%    1.266 & 93.00 & 13.05 & 2.1 & -0.75 & 957 \\
    1.266 & 92.42 & 13.40 & 4.1 & -1.41 & 691 \\
    1.266 & 92.42 & 13.05 & 1.5 & -0.52 & 1135 \\
   \hline   \hline
  \end{tabular} \end{center}
\caption{The Goldberger--Treiman discrepancy $\Delta_{\pi N}$ for
    various input parameters. Also given are the LEC $b_{23}$ and the
    pion--nucleon cut--off as explained in the text.\label{tab:GTR}}
\end{table}

%%%%%%%%%%%%%%%%%%%%%%%%%%%%%%%%%%%%%%%%%%%%%%%%%%%%%%%%%%%%%%%%%%%%%%%%%%%%
\section{Problems, open questions and omissions}
\label{sec:final}
\setcounter{equation}{0}

Here, I will list a few topics which deserve further study. This list
should neither be considered complete nor does the ordering imply any
priority. I also mention a few interesting and important developments
which were not discussed in any detail.

\begin{enumerate}
%\item[(i)] In the two--flavor sector, we have to perform the one--loop
%calculation for $\pi N \to \pi \pi N$ (to order $q^3$ or even better, $q^4$).
%At present, only tree level calculations and the first corrections from
%${\cal L}_{\pi N}^{(2)}$ are available. This is in marked contrast  to
%the many existing rather precise data. What one finally wants to learn
%is to  what {\it precision} these threshold pion production data encode
%information about the low energy elastic $\pi \pi$ scattering
%amplitude. Presently available determimations of the S--wave $\pi \pi$
%scattering lenghts from these data should only be considered as estimates.
%
%\smallskip
%
\item[(i)] More precise data to which HBCHPT can be applied are
needed. This would then allow for a systematic study of the LECs
appearing in ${\cal L}_{\pi N}^{(3,4)}$ and to judge the valitidy
(quality) of the resonance saturation principle which is often used to
get a handle on the LECs. This calls for a joint effort of the
experimenters and the theoreticians. Experimental activities are
under way for double neutral pion production off nucleons, ordinary
and radiative muon capture on protons, charged and neutral pion 
electroproduction off nucleons as well as kaon photoproduction off
protons and deuterium.

\smallskip

\item[(ii)] In the three flavor sector, 
we do not yet have a {\it consistent} picture. 
The problems here are related to the facts that a) the expansion
parameter $M_K / (4 \pi F_K) = 0.4$ is not that small and b) there
are in some channels even subthreshold resonances which make
a direct application of CHPT problematic. Only a few  complete (and
in some cases accurate) calculations exist, i.e. we can not yet
draw decisive conclusions. A status report is given in \cite{ulfsu3}. There
exist many data and more are coming, as an example let me just mention the
accurate threshold kaon photoproduction ones from ELSA or the
proposals to measure the hyperon polarizabilities at Fermilab and
CERN and, of course, the kaon production data off nucleons and
deuterium from Jefferson Lab. More theoretical effort is needed
to clarify the situation.

\smallskip

\item[(iii)] Jenkins and Manohar  first advocated to supplement the EFT
of the ground state baryons and Goldstone bosons by the spin-3/2
decuplet \cite{jmdel} \cite{dobo}. This approach has been taken up in
quite a few papers thereafter. If one thinks of extending calculations
like $\pi N \to \pi N$ through the $\Delta$ region, this is certainly 
unavoidable. Recently, Hemmert et al.\cite{hhk} have developed a
framework to consistently treat the residual  octet--decuplet mass
difference in a systematic fashion extending the path integral
formalism of \cite{BKKM}. This so--called "small scale"
expansion is not exactly a low--energy expansion since in the chiral
limit, the $N\Delta$ splitting remains finite. 
What is missing is a set of {\it complete}
calculations for  observables from which one could  assess
the accuracy of the approach. In particular, it remains to be
seen that in cases where the conventional picture fails, the 
small scale expansion indeed works better. Work along these
lines is underway for Compton scattering~\cite{hhk1}, pion
photoproduction in the $\Delta$ region and other processes.

\smallskip

\item[(iv)] The extension of the CHPT approach to systems  of two or
more nucleons has only begun. Despite some theoretical problems (the
power counting only applies to the subset of irreducible diagrams),
it seems to shed some light on the phenomenology of nuclear forces like
e.g. the smallness of three--, four-- , $\dots$ body forces and the
masking of isospin violation in these systems 
\cite{weinnn}\cite{weinnp}\cite{ubi0}\cite{ubi1}\cite{ubi2}\cite{egm}. 
However, it is
mandatory to perform these calculations with {\it all} the input which
is available from the single baryon sector. This has not yet been
done. Most promising seems to be a mixed approach, in which one
uses CHPT to calculate the kernel of the process under consideration
and sews that together with precise wave functions from
boson--exchange models~\cite{swnp2}. The most precise calculation
so far concerns neutral pion production off deuterium discussed
before. A review is given in \cite{ubi3}.

\smallskip

\item[(v)] Throughout these lectures, I have been rather casual in discussing
the precise relation between S--matrix elements calculated in the heavy
fermion approach and the relativistic version. This has been worked out in
detail in ref.~\cite{emwf}. As an example, consider the tree graphs with
fixed LECs contributing to the threshold parameters in pion--nucleon 
scattering. In ref.~\cite{bkmlec}, these were calculated by working out the
{\it relativistic} Born graphs and then expanding the resulting expressions
in powers of $M_\pi / m$. On the
other hand, the same calculation in ref.~\cite{moj} was performed entirely in
the heavy baryon framework with no recourse to the relativistic theory.
Of course, these approaches have to be completely equivalent. This can 
be seen if one studies carefully the wave function renormalization in
HBCHPT, which shows some intriguing features. For details, I refer to
\cite{emwf}. For purely calculational purposes, I recommend the procedure
used e.g. in \cite{bkmpi0} or \cite{bkmlec}.

\end{enumerate}

\bigskip
%%%%%%%%%%%%%%%%%%%%%%%%%%%%%%%%%%%%%%%%%%%%%%%%%%%%%%%%%%%%%%%%%%%%%
\section*{Acknowledgements}
\addcontentsline{toc}{section}{\numberline{}Acknowledgements}
First, I would like to thank the organizers,
in particular Jos\'e Goity, for their kind invitation
and hospitality. I am also grateful to  V\'{e}ronique
Bernard, Norbert Kaiser, Guido M\"uller and Sven Steininger
for fruitful collaborations and allowing me
to present some material before publication.

\end{document}